\RequirePackage{fix-cm}
\documentclass[smallextended]{svjour3}       
\smartqed  
\usepackage{appendix}
\usepackage{amsmath}
\usepackage{graphicx}
\usepackage{array}
\usepackage{longtable}

\usepackage{natbib}

\usepackage[graphicx]{realboxes}
\usepackage[table,xcdraw]{xcolor}
\usepackage{tablefootnote}
\usepackage{hyperref}
\usepackage{color}
\usepackage[table,xcdraw]{xcolor}
\usepackage[graphicx]{realboxes}
\usepackage{multirow}

\usepackage{lipsum}

\usepackage{float} 

\usepackage{breqn} 

\usepackage{url}

\usepackage[most]{tcolorbox}

\usepackage{amssymb}

\usepackage{enumitem}
\usepackage{tcolorbox} 
\usepackage{textcomp} 

\definecolor{hzcolor}{RGB}{10, 186, 181}

\definecolor{atcolor}{RGB}{200, 25, 130}

\definecolor{summary_color}{rgb}{0.7, 0.75, 0.71}

\newtcolorbox[auto counter]{summary}[1][]{title={\bfseries Summary of RQ~\thetcbcounter},enhanced,drop shadow={black!50!white},
  coltitle=black,
  top=0.3in,
  attach boxed title to top left=
  {xshift=1.5em,yshift=-\tcboxedtitleheight/2},
  boxed title style={size=small,colback=pink},#1}

\newtcolorbox[]{summary_new}[2][]{title={#2},enhanced,drop shadow={black!50!white},
  coltitle=black,
  top=0.3in,
  attach boxed title to top left=
  {xshift=1.5em,yshift=-\tcboxedtitleheight/2},
  boxed title style={size=small,colback=pink},#1}

\definecolor{gopicolor}{RGB}{40,50 , 181}

\begin{document}

\sloppy

\title{Revisiting reopened bugs in open source software systems
}


\author{Ankur~Tagra         \and
        Haoxiang~Zhang \and Gopi~Krishnan~Rajbahadur \and Ahmed~E.~Hassan 
}


\institute{Ankur Tagra \and Ahmed E. Hassan \at
              {Software Analysis and Intelligence Lab (SAIL) \\
              Queen's University\\
              Kingston, ON, Canada}\\
              \email{\{atagra, ahmed\}@cs.queensu.ca}\\
              \and Haoxiang Zhang \and Gopi~Krishnan~Rajbahadur \at
              {Centre for Software Excellence at Huawei, Canada}\\
              \email{haoxiang.zhang@acm.org, gopi.krishnan.rajbahadur1@huawei.com}\\
              \\
}

\date{Received: date / Accepted: date}

\maketitle

\begin{abstract}
Reopened bugs can degrade the overall quality of a software system since they require unnecessary rework by developers. Moreover, reopened bugs also lead to a loss of trust in the end-users regarding the quality of the software. Thus, predicting bugs that might be reopened could be extremely helpful for software developers to avoid rework. Prior studies on reopened bug prediction focus only on three open source projects (i.e., Apache, Eclipse, and OpenOffice) to generate insights. We observe that one out of the three projects (i.e., Apache) has a data leak issue -- the bug status of \textit{reopened} was included as training data to predict reopened bugs. In addition, prior studies used an outdated prediction model pipeline (i.e., with old techniques for constructing a prediction model) to predict reopened bugs. Therefore, we revisit the reopened bugs study on a large scale dataset consisting of 47 projects tracked by JIRA using the modern techniques such as SMOTE, permutation importance together with 7 different machine learning models. We study the reopened bugs using a mixed methods approach (i.e., both quantitative and qualitative study). We find that: 1) After using an updated reopened bug prediction model pipeline, only 34\% projects give an acceptable performance with AUC $\geqslant$ 0.7. 2) There are four major reasons for a bug getting reopened, that is, technical (i.e., patch/integration issues), documentation, human (i.e., due to incorrect bug assessment), and reasons not shown in the bug reports. 3) In projects with an acceptable AUC, 94\% of the reopened bugs are due to patch issues (i.e., the usage of an incorrect patch) identified before bug reopening. Our study revisits reopened bugs and provides new insights into developer's bug reopening activities.

\end{abstract}

\section{Introduction}
\label{intro}
\begin{figure}[!ht]
\centering
  \includegraphics[width=0.8\textwidth]{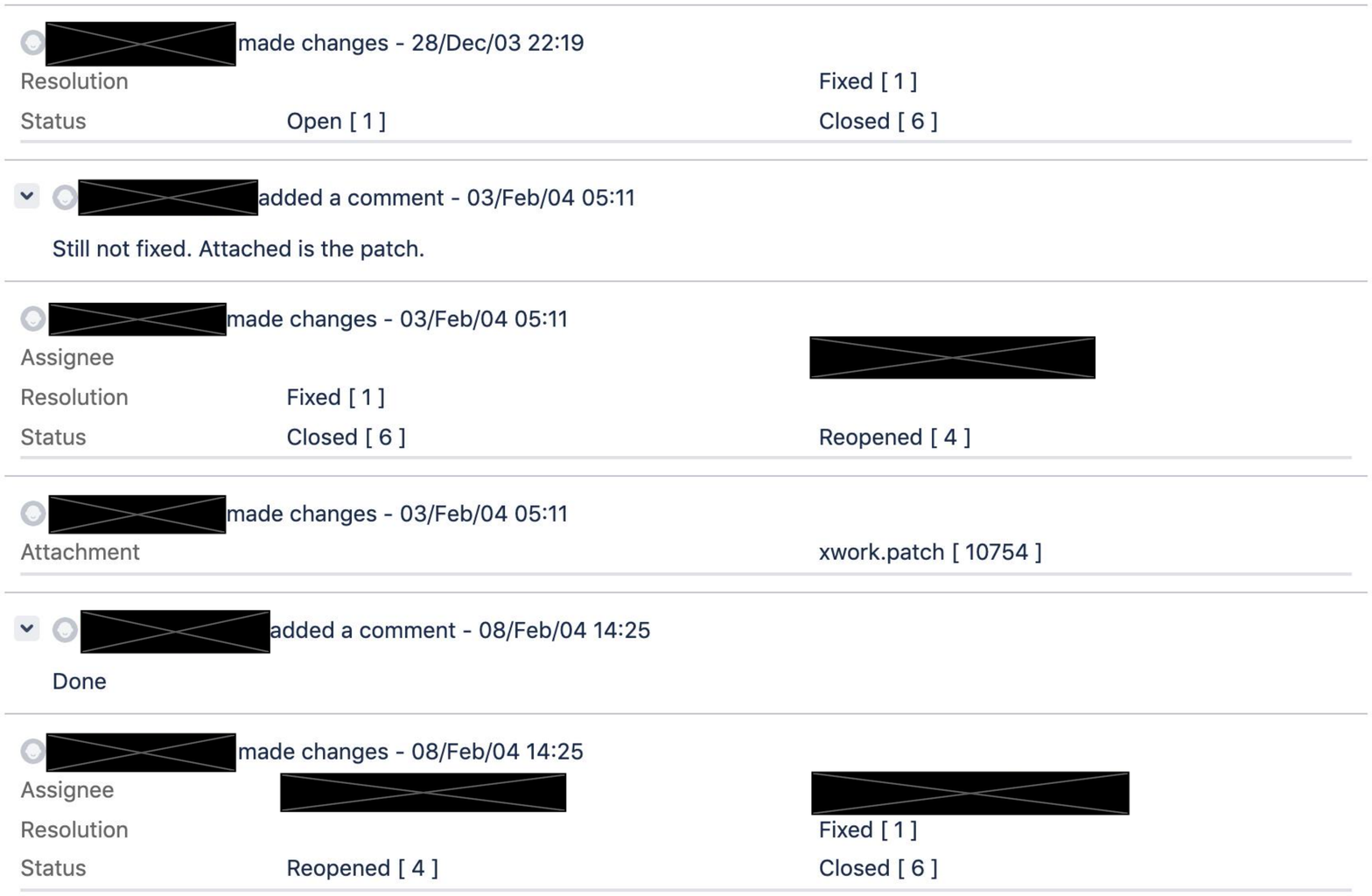}
\caption{Historical events in a reopened bug}
\label{fig_sample_reopened_bug_report_comments}       
\end{figure}

Fixing bugs is an important part of software development. Generally, a bug is resolved and closed after the bug is fixed. However, sometimes a bug needs to be reopened. 
\cite{mi2016empirical} observed that around 6--10\% of the bugs are reopened in four open source projects from the Eclipse product family. If a fair number of fixed bugs get reopened, it is an indication of instability in the software system \citep{zimmermann2012characterizing}. Reopened bugs consume more time to resolve than normal bugs by a factor of 1.6--2.1 \citep{mi2016empirical}. Reopened bugs cause extra rework effort for development teams \citep{mi2018not}. Moreover, reopened bugs also lead to a loss of trust in the end-users regarding the quality of the software \citep{shihab2010predicting}. Figure \ref{fig_sample_reopened_bug_report_comments} shows a reopened bug\footnote{\url{https://issues.apache.org/jira/browse/XW-140}}. In this issue, the bug was \textit{fixed} and \textit{closed}; however, later another developer reopened the bug and fixed it. It took the bug an extra 42 days from initial \textit{close} to the bug getting \textit{reopened} and finally \textit{closed}.

Therefore, to avoid the reopening of a bug, prior studies proposed models to predict if a bug will be reopened (i.e., reopened bug prediction models). For example, \cite{shihab2013studying} studied the reopening likelihood of a bug based on four dimensions of features (i.e. work habits, bug report, bug fix, and team dimension). They studied three projects (i.e. Apache, Eclipse, and OpenOffice) using 44,317 non-reopened bugs and 11,745 reopened bugs. \cite{xia2015automatic} proposed the ReopenPredictor\footnote{https://github.com/xin-xia/reopenBug} and further improved the performance of the reopened bug prediction model using the same dataset as used by Shihab et al. \citep{shihab2013studying}. Xia et al. \citep{xia2015automatic} found that the model to predict reopened bugs show a consistent performance on the three studied projects. We examined the bug reports from the data shared by Shihab et al. \citep{shihab2013studying} and Xia et al. \citep{xia2015automatic}, we observed that one out of the three projects (i.e., Apache project) has a data leak issue \citep{kaufman2012leakage, tu2018careful}. Several instances of the feature \textit{last status} contained the value of reopened (not all the instances of Apache project). This suggests that the dataset to train the prediction model of reopened bugs contained the information that a bug was reopened already. When the data from the future is used to predict current events, it amounts to data leak \citep{kaufman2012leakage}. Therefore, if we discount the findings from the Apache project, the findings of the prior studies are effectively based on only two projects (i.e., Eclipse, and OpenOffice). Furthermore, all these prior studies used an old prediction pipeline and with time, the techniques for prediction models have improved. For instance, no prior study in reopened bug prediction tuned the hyperparameters of their studied model, which could, in turn, improve the prediction performance of the constructed reopened bug prediction model \citep{rajbahadur2019impact, tantithamthavorn2018impact}. No prior study in reopened bug prediction performed correlation and redundancy analysis on the independent features of their dataset to remove correlated features; however, several recent studies have used correlation and redundancy analysis on their dataset prior to training the prediction model \citep{jiarpakdee2019impact,tantithamthavorn2018impact,rajbahadur2019impact,rajbahadur2017impact,lee2020empirical, da2018impact, mcintosh2016empirical}. This motivated us to do a large scale study of the likelihood of reopening a bug using the latest prediction pipeline. In our study, we answer the following research questions.

\paragraph{\normalfont \textbf{RQ1: How well do bug reopen models generalize?\\}}
\parshape 
  2
  0pt \linewidth
  \parindent \dimexpr\linewidth-\parindent
\textbf{\textit{Motivation:}} Prior studies on reopened bugs focus only on three projects (Apache, Eclipse, OpenOffice) \citep{shihab2013studying, xia2015automatic}. Out of these three projects Apache project has data leak issues. In addition, Several components used for the prediction of reopened bugs (such as the strategy for handling class imbalance, generating feature importance of the prediction model) have shown to be not as effective (e.g., the strategy for handling class imbalance, generating feature importance of the prediction model). Therefore, in this RQ, we revisit the prior findings of reopened bug prediction on a large scale study of 47 new projects that we collect from JIRA as we outline in Section \ref{data} and with the latest techniques for building reopened bug prediction pipeline to determine the likelihood of bug reopening.\\
\textbf{\textit{Results:}} Using our updated reopened bug prediction model pipeline, we observe that across the 47 studied projects, at best we are able to achieve an acceptable AUC ($\geqslant$ 0.7) on 34\% (16/47) of the projects. Furthermore, reopened bug prediction models constructed with Random Forest and Gradient Boosting models yield an acceptable AUC ($\geqslant$ 0.7) on more studied projects than the model constructed with other classifiers (for example, Decision Tree and Adaboost). Finally, with our large scale study, we observe that \textit{number of comments} is the most important feature that helps in predicting the reopening likelihood of a bug. Such a finding contrasts with the findings observed by Shihab et al. \citep{shihab2013studying} where they found that the \textit{comment text} is the most important feature in determining the likelihood of reopening a bug.\\



\paragraph{\normalfont \textbf{RQ2: Why do bugs get reopened?\\}}
\parshape 
  2
  0pt \linewidth
  \parindent \dimexpr\linewidth-\parindent
\textbf{\textit{Motivation:}} \cite{zimmermann2012characterizing} observe that understanding bug reopening is required since it helps in various software development tasks such as identifying areas which need better tool support and improving the bug triage process. We observe by interpreting the reopened bug prediction model that \textit{number of comments} and \textit{comment text} are the top two most important features in determining the likelihood of reopening a bug. These insights are coarse and do not provide enough actionable suggestions that can be used to either prevent bug reopening or identify the potential root cause of reopened bugs. Therefore, in this RQ, we investigate the rationale behind why bugs are reopened by leveraging the whole history of bug reports. In addition, we investigate if the reasons for reopening a bug differ between the projects on which the constructed reopened bug prediction models have an acceptable and a poor performance.\\
\textbf{\textit{Results:}} Through a mixed method approach, we identify that in 63\% and 86\% of the reopened bugs, developers post comments during and after the reopening respectively, and in 77\% of the reopened bugs, the reason to reopen a bug is identified either during or after the bug reopening. Through a qualitative study of 370 samples of reopened bugs (with a 95\% confidence level and a 5\% confidence interval), we identify four categories of reasons for reopening a bug, that is, technical (i.e., patch/integration issues), documentation (i.e., updating fields of a bug report), human (i.e., incorrect assessment by a developer), and no stated reason in a bug report. In addition, through our qualitative study, we observe that in projects with an acceptable AUC, 94\% of the reopened bugs are due to patch issues and a negligible proportion of 4\% bugs are reopened due to integration issues, whereas in projects with a poor AUC, apart from patch issues (66\%), bugs are reopened due to integration issues (22\%).\\

Based on our findings, we encourage developers to carefully examine their patches for bug fixing, e.g., by testing the patches thoroughly before resolving the bugs, as the most common reasons for reopening bugs are due to technical issues (i.e., patch/integration issues). Also, we suggest that the bug tracking systems should enable developers to edit certain fields of bug reports without requiring the reopening of a bug, as 24\% of our manually studied reopened bugs are reopened due to documentation (bookkeeping) issues. 


The rest of the paper is organized as follows. Section \ref{data} describes the data collection processes. Section \ref{Methodology} discusses our case study.  Section \ref{Discussion} presents the implications of our findings. Section \ref{RelatedWork} discusses the related work. Section \ref{threats} provides the threats to the validity of our study. Finally, Section \ref{Conclusion} discusses the conclusion of our study.





\section{Data Collection and Preprocessing}
\label{data}
We describe our data collection process in this section. For revisiting the reopened bugs study on a large dataset, we collect 47 projects from JIRA (i.e., JIRA dataset) and map all the features of the Bugzilla dataset that are also present in the JIRA dataset. In Section \ref{dataset_collection}, we describe the data collection processes of the JIRA dataset.
The JIRA dataset contains categorical (i.e., nominal), numeric, boolean, and text features. We discuss these features in detail in Section \ref{dataset_collection},
and in Section \ref{subsection:preprocessing} we discuss our process to clean the data for the reopened bug models.

\subsection{Collecting the JIRA dataset}
\label{dataset_collection}

We study the reopened bugs across all public projects that are tracked by JIRA and whose source code is hosted on Github. JIRA is a commonly used bug tracking system and it is custom tailored to support 3,000 apps\footnote{\url{https://www.atlassian.com/software/jira}}. We use the JIRA REST API\footnote{\url{https://developer.atlassian.com/cloud/jira/platform/rest/v2/intro/}} to extract JIRA bugs. We use query subfield \textit{type=bug} to ensure that we only collect the bugs from JIRA. Firstly, we collected all the bugs that were ever reopened across all public projects tracked by Apache JIRA. For non-reopened bugs, we collected all the bugs which were either closed or resolved but were never reopened. In total, we collected 388,404 non-reopened bugs and 25,376 reopened bugs across 589 projects. All the features except the \textit{number of files in the fix} were collected using the JIRA REST API. We used Pydriller\footnote{\url{https://pydriller.readthedocs.io/en/latest/}} to collect the number of files changed in the corresponding GitHub repositories using the approach outlined by \cite{vieira2019reports}. Table~\ref{table:feature_shihabvs_our} shows the 20 features that we collect for our study. Note that the features (i.e., \textit{platform}, \textit{severity}, \textit{severity changed} and \textit{number in the CC list}) are not present in the JIRA. We further filter the studied projects using two criteria: 1) we select the top 50 projects in JIRA based on the committers count\footnote{\url{https://projects.apache.org/projects.html?number}} since they are quite representative of popular Apache projects, 2) We remove the projects that do not have at least 10 reopened bugs. We do so, as in Section \ref{RQ2}, we discuss that, in our model we use 10-times-stratified 10-fold cross validation. When we 10-fold split a dataset with fewer than 10 samples of positive class (reopened bugs), there will be at least one fold with no positive samples. Thus the model will not get trained for that fold (unavailability of positive sample). By applying these two criteria, we end up with 47 projects to study (including 178,636 non-reopened bugs and 9,993 reopened bugs). We provide the replication package\footnote{\url{https://www.dropbox.com/sh/mb4759od1p1oheu/AAB84v1DOJ1R6JFJU92Nyo5ha?dl=0}} with data of the 47 projects in our study. While computing the value of features, we ignore all events in the bug report that occur at the time of reopening the bug or after reopening the bug, to prevent any data leak issue. For example, for the \textit{comment text} feature, we consider all the comments before the first reopening of each studied bug. We do not consider comments of subsequent reopens in a bug report. A bug can be reopened multiple times; however, in this study we only study the prediction of the first reopening of a bug by leveraging activities before the first reopening of a bug.

\begin{table}[]
\caption{The features that we use in our reopened bug prediction model}
\label{table:feature_shihabvs_our}
\begin{tabular}{|l|l|l|l|l|}
\hline
\textbf{Category}            & \textbf{\begin{tabular}[c]{@{}l@{}}Features\\ used by\\ Shihab et al.\end{tabular}} & \textbf{\begin{tabular}[c]{@{}l@{}}Our \\ features\\  (JIRA)\end{tabular}} & \textbf{Type} & \textbf{Explanation}                                                                                                              \\ \hline
\multirow{5}{*}{Work habits} & Time                                                                                & Time                                                                       & Nominal       & \begin{tabular}[c]{@{}l@{}}Time of the day when\\first closing the bug\\ (e.g., morning)\end{tabular} \\ \cline{2-5} 
                             & Weekday                                                                             & Weekday                                                                    & Nominal       & \begin{tabular}[c]{@{}l@{}}Day of the week when\\ first closing the bug\\ (e.g., monday)\end{tabular}                              \\ \cline{2-5} 
                             & \begin{tabular}[c]{@{}l@{}}Month\\ day\end{tabular}                                 & \begin{tabular}[c]{@{}l@{}}Month\\ day\end{tabular}                        & Numeric       & \begin{tabular}[c]{@{}l@{}}Day of the month when\\ first closing the bug\end{tabular}                             \\ \cline{2-5} 
                             & Month                                                                               & Month                                                                      & Numeric       & \begin{tabular}[c]{@{}l@{}}Month when first closing\\ the bug\end{tabular}                                          \\ \cline{2-5} 
                             & \begin{tabular}[c]{@{}l@{}}Day of\\ year\end{tabular}                               & \begin{tabular}[c]{@{}l@{}}Day of\\ year\end{tabular}                      & Numeric       & \begin{tabular}[c]{@{}l@{}}Day of the year when first\\ closing the bug\end{tabular}                                  \\ \hline
\multirow{12}{*}{Bug report} & Component                                                                           & Component                                                                  & Nominal       & \begin{tabular}[c]{@{}l@{}}The component of the\\ project that the bug\\ belong to\end{tabular}                                                               \\ \cline{2-5} 
                             & Platform                                                                            & -                                                                          & -             & -                                                                                                                                 \\ \cline{2-5} 
                             & Severity                                                                            & -                                                                          & -             & -                                                                                                                                 \\ \cline{2-5} 
                             & Priority                                                                            & Priority                                                                   & Numeric       & The priority of the bug                                                                                                          \\ \cline{2-5} 
                             & \begin{tabular}[c]{@{}l@{}}Number in\\ the CC list\end{tabular}                     & -                                                                          & -             & -                                                                                                                                 \\ \cline{2-5} 
                             & \begin{tabular}[c]{@{}l@{}}Description\\ size\end{tabular}                          & \begin{tabular}[c]{@{}l@{}}Description\\ size\end{tabular}                 & Numeric       & \begin{tabular}[c]{@{}l@{}}The number of words in\\ the description\end{tabular}                                                 \\ \cline{2-5} 
                             & \begin{tabular}[c]{@{}l@{}}Description\\ text\end{tabular}                          & \begin{tabular}[c]{@{}l@{}}Description\\ text\end{tabular}                 & Text          & \begin{tabular}[c]{@{}l@{}}The text of the\\ description of a bug report\end{tabular}                                                            \\ \cline{2-5} 
                             & \begin{tabular}[c]{@{}l@{}}Number of\\ comments\end{tabular}                        & \begin{tabular}[c]{@{}l@{}}Number of\\ comments\end{tabular}               & Numeric       & \begin{tabular}[c]{@{}l@{}}The number of comments\\ in a bug report\end{tabular}                                                      \\ \cline{2-5} 
                             & \begin{tabular}[c]{@{}l@{}}Comment\\ size\end{tabular}                              & \begin{tabular}[c]{@{}l@{}}Comment\\ size\end{tabular}                     & Numeric       & \begin{tabular}[c]{@{}l@{}}The number of words in\\ the comments\end{tabular}                                                    \\ \cline{2-5} 
                             & \begin{tabular}[c]{@{}l@{}}Comment\\ text\end{tabular}                              & \begin{tabular}[c]{@{}l@{}}Comment\\ text\end{tabular}                     & Text          & \begin{tabular}[c]{@{}l@{}}The text of the comments\\ from a bug report\end{tabular}                                                                                                \\ \cline{2-5} 
                             & \begin{tabular}[c]{@{}l@{}}Priority\\ changed\end{tabular}                          & \begin{tabular}[c]{@{}l@{}}Priority\\ changed\end{tabular}                 & Boolean       & \begin{tabular}[c]{@{}l@{}} Whether the priority level \\changed in a bug report\end{tabular}                                                                                                        \\ \cline{2-5} 
                             & \begin{tabular}[c]{@{}l@{}}Severity\\ changed\end{tabular}                          & -                                                                          & -             & -                                                                                                                                 \\ \hline
\multirow{3}{*}{Bug fix}     & Time days                                                                           & Time days                                                                  & Numeric       & \begin{tabular}[c]{@{}l@{}}The time it took to close\\ the bug\end{tabular}                                                      \\ \cline{2-5} 
                             & \begin{tabular}[c]{@{}l@{}}Last \\ status\end{tabular}                              & \begin{tabular}[c]{@{}l@{}}Last\\ status\end{tabular}                      & Nominal       & \begin{tabular}[c]{@{}l@{}}The last status when the bug\\ was first closed\end{tabular}                                          \\ \cline{2-5} 
                             & \begin{tabular}[c]{@{}l@{}}Number of\\ files in the\\ fix\end{tabular}              & \begin{tabular}[c]{@{}l@{}}Number of\\ files in the\\ fix\end{tabular}     & Numeric       & \begin{tabular}[c]{@{}l@{}}The number of files in the\\ bug fix\end{tabular}                                                         \\ \hline
\multirow{4}{*}{People}      & \begin{tabular}[c]{@{}l@{}}Reporter\\ name\end{tabular}                             & \begin{tabular}[c]{@{}l@{}}Reporter\\ name\end{tabular}                    & Nominal       & \begin{tabular}[c]{@{}l@{}}The name of the reporter\\ in the  bug report\end{tabular}                                                    \\ \cline{2-5} 
                             & \begin{tabular}[c]{@{}l@{}}Fixer \\ name\end{tabular}                               & \begin{tabular}[c]{@{}l@{}}Assignee\\ name\end{tabular}                    & Nominal       & \begin{tabular}[c]{@{}l@{}}The name of the assignee\\ in the bug report\end{tabular}                                                    \\ \cline{2-5} 
                             & \begin{tabular}[c]{@{}l@{}}Reporter\\ experience\end{tabular}                       & \begin{tabular}[c]{@{}l@{}}Reporter\\ experience\end{tabular}              & Numeric       & \begin{tabular}[c]{@{}l@{}}The number of bugs that\\ a reporter has reported\\ before reporting this bug\end{tabular}          \\ \cline{2-5} 
                             & \begin{tabular}[c]{@{}l@{}}Fixer \\ experience\end{tabular}                         & \begin{tabular}[c]{@{}l@{}}Assignee\\ experience\end{tabular}              & Numeric       & \begin{tabular}[c]{@{}l@{}}The number of bugs\\ assigned to the assignee\\ before this bug\end{tabular}                          \\ \hline
\end{tabular}
\end{table}


\subsection{Pre-processing features}
\label{subsection:preprocessing}

\begin{figure}[!ht]
\centering
  \includegraphics[width=0.7\textwidth]{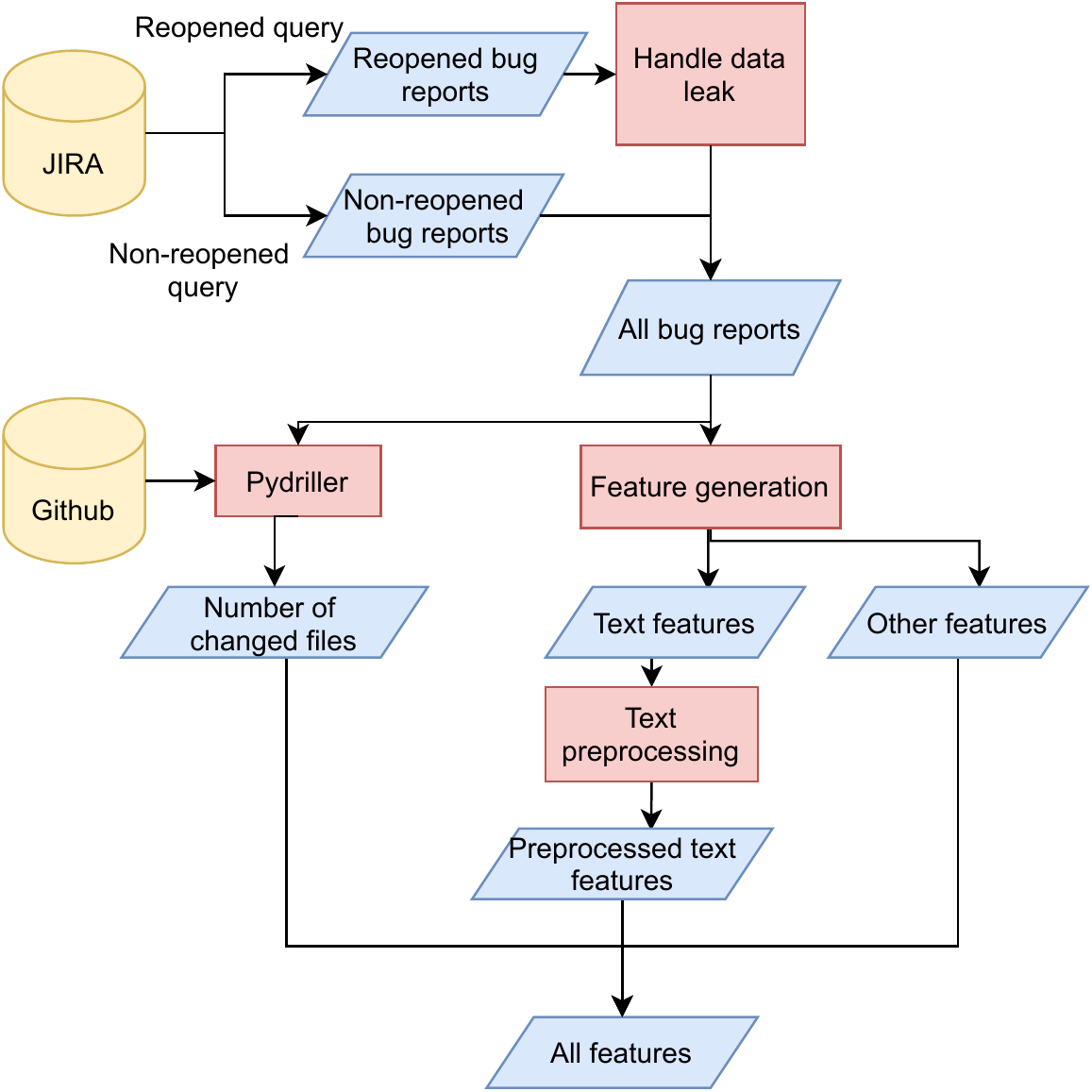}
\caption{Workflow of data collection and pre-processing}
\label{fig_data_collection}       
\end{figure}

Figure \ref{fig_data_collection} shows the steps we followed to collect bug reports and how we process the data. Our dataset contains two text features, i.e., the \textit{description text} and \textit{comment text}. We pre-processed the text features similar to prior studies~\citep{uysal2014impact, denny2017text, kannan2014preprocessing}. The comments and description of a bug report contain URLs. We replace them with a placeholder token to simplify the representation of URLs~\citep{hemalatha2012preprocessing}. Developers add code blocks in the description or comments of a bug report. To simplify our text pre-processing, we replace code blocks with a placeholder token. Error logs and stack traces are processed similarly, i.e., by replacing them with a placeholder token. We then remove all the stop words as they do not contribute much to the context of reopened bugs \citep{song2005comparative}. We then use the NLTK package\footnote{\url{https://www.nltk.org/}} to stem the text to reduce the different forms of a word \citep{srividhya2010evaluating, mendez2005tokenising, hardeniya2016natural}. Finally, we use the pre-processed text in our reopened bug prediction model together with the other features that are categorical (e.g., \textit{component}, \textit{last status}, and \textit{assignee}), numeric (e.g., \textit{month}, \textit{description size}, and \textit{number of comments}), or boolean (e.g., \textit{priority changed}).

\section{Case Study}
\label{Methodology}
In this section, we describe our case study results based on our new dataset to study reopened bugs. We first use the new JIRA dataset to predict reopened bugs (in Section \ref{RQ2}). Then we analyze the studied bugs to understand the rationale for reopening a bug report (in Section \ref{RQ3}). For each RQ, we provide the motivation, approach, and results of the RQ.

\subsection{RQ1: How well do bug reopen models generalize?}
\label{RQ2}
\textbf{Motivation:}
Prior studies on determining the likelihood of reopening a bug by \cite{shihab2013studying} and later by~\cite{xia2015automatic} were based only on three projects (i.e., Apache, Eclipse, and OpenOffice). Moreover, the other two studies by \cite{shihab2010predicting} and by \cite{xia2013comparative} on reopened bug prediction were based only on one project (i.e., Eclipse). After examining the dataset used by these prior studies, we observe that out of their studied three projects, one project (i.e., Apache) has a data leak issue -- the bug status of \textit{reopened} was included as training data to predict reopened bugs. Hence, prior findings on determining the likelihood of reopening a bug were effectively based only on two projects. In addition, these prior studies were conducted 6--10 years ago. Several experimental components that the prior studies used to build a reopened bug prediction model pipeline have since then been shown to be not as effective when building a prediction model. For example, prior studies on reopened bug prediction did not tune the hyperparameters of their used model. However, several recent studies have shown that tuning the hyperparameters of a model is pivotal to ensure its optimal performance and interpretation~\citep{fu2016tuning,tantithamthavorn2018impact,tantithamthavorn2016automated}. Furthermore, prior studies \citep{shihab2013studying} used a class re-weighting and re-sampling strategy to deal with the class imbalance issue that is present in the projects. However, several recent studies have shown SMOTE (Synthetic Minority Over-sampling Technique) to be a more effective method for class re-balancing~\citep{tantithamthavorn2018impact,agrawal2018better}. Prior studies \citep{shihab2013studying} used top node analysis with a Decision Tree model for generating feature importance of their model; however, recent studies show that the results of top node analysis can be biased since Decision Tree favors categorical features \citep{altmann2010permutation}. Moreover, prior studies \citep{shihab2013studying, xia2015automatic} did not use feature encoding with the latest and more widely used encoding techniques \citep{cerda2018similarity} such as one-hot encoding to encode the categorical features. 

After Shihab et al. \citep{shihab2013studying} and Xia et al.'s \citep{xia2015automatic} study, no subsequent study revisited the reopened bug prediction problem with these pipeline improvements in mind. Therefore, in this RQ, we revisit their findings on a large scale study of 47 new projects that we collect from JIRA as we outline in Section \ref{data} and with the latest techniques for building reopened bug prediction model pipeline to determine the likelihood of bug reopening.\\

\begin{figure}[!ht]
\centering
  \includegraphics[width=0.7\columnwidth]{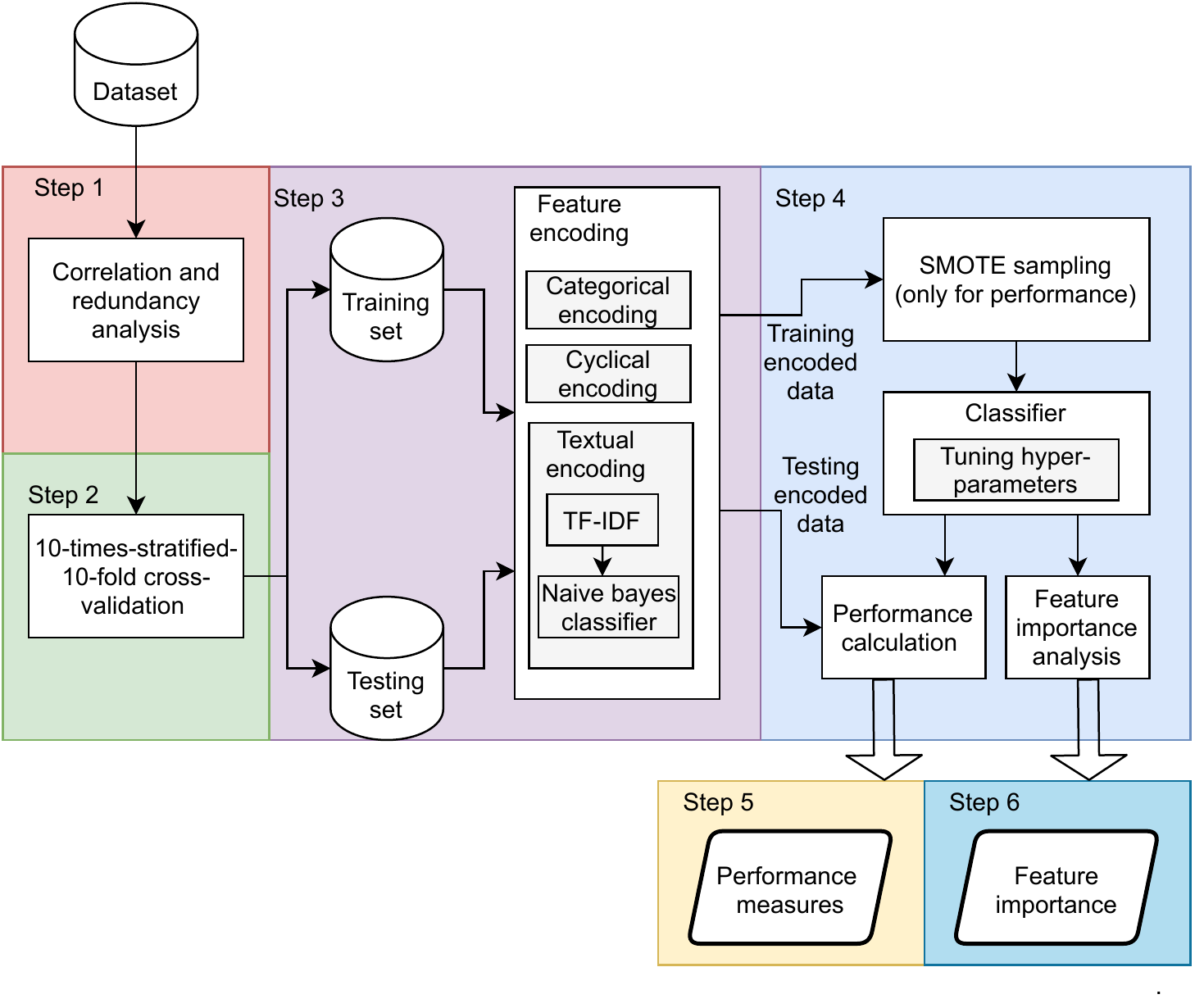}
\caption{Reopened bug prediction model pipeline}
\label{fig_RQ1_approach}       
\end{figure}

\noindent \textbf{Approach:}
We construct our reopened bug prediction model in this RQ. Figure \ref{fig_RQ1_approach} shows the prediction model pipeline that we use to construct our reopened bug prediction model. We discuss each step of the pipeline in detail below.

\begin{quote}
\noindent\textbf{Step 1: Correlation and redundancy analysis.} \cite{tantithamthavorn2018impact} outlines that presence of correlated and redundant features might result in affecting the computed feature importance ranks of a model. Therefore, similar to prior studies \citep{jiarpakdee2019impact,tantithamthavorn2018impact,rajbahadur2019impact,rajbahadur2017impact,lee2020empirical, da2018impact, mcintosh2016empirical}, we perform a correlation and redundancy analysis on the independent features of our dataset to remove correlated features using the implementation provided by AutoSpearman method of Rnalytica\footnote{\url{https://rdrr.io/github/software-analytics/Rnalytica/man/AutoSpearman.html}} R package \citep{tantithamthavorn2018impact, lee2020empirical,yatish2019mining}. We choose the AutoSpearman method to remove the correlated and redundant features as the study by \cite{jiarpakdee2019impact} showed that selecting features with AutoSpearman typically yields better subset of features than other techniques. In addition, Jirapakdee et al. also show that when using AutoSpearman, the impact of the feature selection on the performance of the trained model is minimal. Finally, AutoSpearman is a method is commonly used in prior studies \citep{jiarpakdee2019impact, lee2020empirical, rajbahadur2021impact}. Autospearman uses a criteria that selects one feature from a group of highest correlated features which shares the least correlation with other features that are not in the group \citep{jiarpakdee2018autospearman} based on Spearman correlation score. Following which, AutoSpearman then removes the variables which have VIF$>=$5. We use a default value of $\rho$ for the Spearman correlation coefficient threshold to remove the correlated features. 

\noindent\textbf{Step 2: Splitting dataset into training/testing sets.} 
We split the dataset into training and testing sets using 10-times-stratified 10-fold cross-validation as opposed to 10-times 10-fold cross-validation used by Shihab et al. \citep{shihab2013studying}. In stratified 10-fold cross-validation, the whole dataset is split into 10 folds in such a way that each fold contains the same percentage of all class labels as in the overall dataset. Each fold is then used to test the model performance and the remaining 9 folds are used for training the model. We repeat this process 10 times. Therefore, for each iteration 10 performance and feature importance measures are generated and an overall 100 performance and feature importance measures are generated on each studied project. For more details of the stratified k-fold cross-validation, please refer to \cite{zeng2019automatic}. We choose 10-times-stratified-10-fold cross-validation in particular because our studied dataset is imbalanced, and stratified k-fold cross-validation helps preserve the original class distribution of the data \citep{he2013imbalanced,arellano2019epidemiological}.

\noindent\textbf{Step 3: Feature encoding.} 

\noindent\textbf{Step 3.1 Categorical features encoding.} 
We encode the categorical features in the dataset using one-hot feature encoding to transform categorical features present in the dataset into numeric features to be used as an input to train the model \citep{zheng2018feature}. We employ one-hot feature encoding as opposed to simply converting the categorical features to numeric features as Shihab et al.~\cite{cerda2018similarity} stated that one-hot feature encoding is one of the latest and more widely used categorical feature encoding procedure. One-hot encoding creates a separate boolean column for each category of the categorical features. These columns are then passed along with the other numeric columns to the next step.

\noindent\textbf{Step 3.2 Cyclical features encoding.} For the feature \textit{weekday} we use cyclical encoding as feature \textit{weekday} values are cyclic in nature~\citep{hebert2020estimation}. We do so as we need to maintain unit distance between each pair of consecutive days for cyclic values \citep{chakraborty2019advanced}. 

\noindent\textbf{Step 3.3 Textual features encoding.} We use the TF-IDF (term frequency-inverse document frequency) vectorization \citep{nyamawe2020feature} to encode the two text features (i.e., \textit{comment text} and \textit{description text}) that are present in the studied datasets similar to several prior studies \citep{biggers2014configuring,mcmillan2011exemplar,rakha2017revisiting,nyamawe2020feature}. TF-IDF works by assigning a higher weight to a word if that word appears many times within a small number of documents. Similarly, TF-IDF assigns lower weight to a word, when the given word appears fewer times in a document. TF-IDF assigns the lowest weight to a word if that given word appears in almost all the documents \citep{corazza2016weighing}. We use the training data in each iteration to build our TF-IDF model and use the trained TF-IDF model to generate the TF-IDF scores for text features in the testing data. 

After transforming the text features with TF-IDF, we encode the text features as probabilities with a Naive Bayes model similar to the prior study \citep{shihab2013studying,meyer2004spambayes,haoxiang_paper_new}. We follow the above approach since using only TF-IDF creates a column in the dataset for each word present in the dataset. Such an encoding might drastically increase the number of studied features and cause our reopened bug prediction models to overfit the studied datasets. Furthermore, when we interpret the constructed reopened bug prediction models, we are more interested in observing the importance of the studied text features (i.e., the \textit{comment text} and \textit{description text}) in predicting if a bug would be reopened (rather than the individual words). Therefore, we train two Naive Bayes models on the training data.

\noindent\textbf{Step 4: Model construction.} 

\noindent\textbf{Step 4.1 Handling class imbalance for the reopened bug prediction model.} 
We show the distribution of percentage of reopened bugs in all projects in Figure \ref{fig_RQ1_percentage}. Since the dataset for determining the likelihood of reopening a bug is imbalanced (i.e., fewer reopened bugs as compared to non-reopened bugs), we used the SMOTE technique to rebalance the training dataset \citep{malhotra2017empirical}. SMOTE is an oversampling approach in which the minority class is over-sampled by generating synthetic samples \citep{chawla2002smote}. We use default values such as \textit{auto} and 5 for parameters sampling strategy and k_neighbors respectively used in SMOTE provided by library imblearn\footnote{\url{https://imbalanced-learn.org/stable/references/generated/imblearn.over_sampling.SMOTE.html}}. \textit{Sampling strategy} is set based on input data type. If \textit{auto} input is provided, it automatically detects the input type and sets the \textit{sampling strategy} accordingly. Shihab et al. \citep{shihab2013studying} used a class re-weighting and re-sampling strategy to deal with the class imbalance present in the dataset. However, we choose SMOTE for handling class imbalance as recent studies show that SMOTE is a more effective method for class re-balancing~\citep{tantithamthavorn2018impact,agrawal2018better}.
\begin{figure}[!ht]
\centering
  \includegraphics[width=0.8\columnwidth]{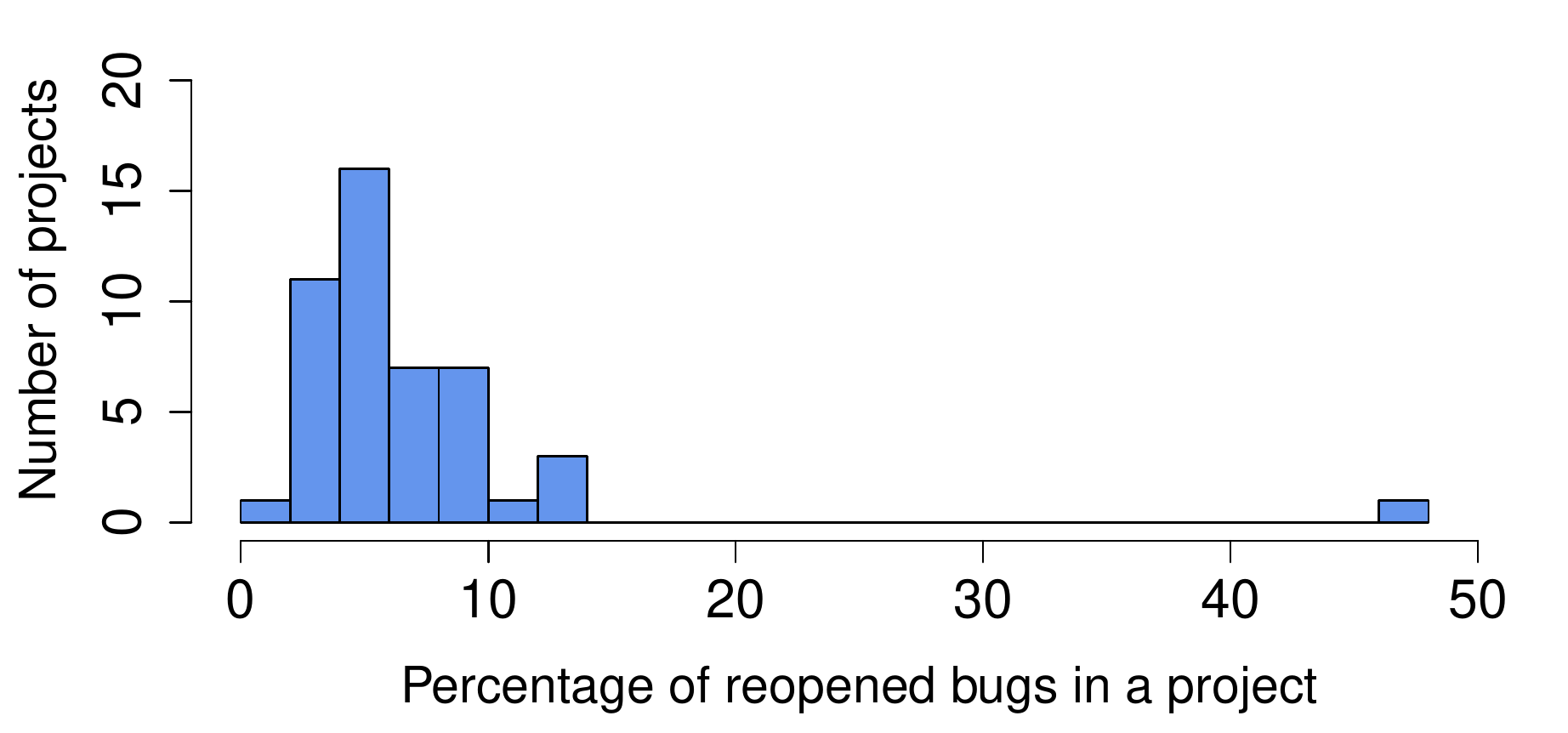}
\caption{Percentage of reopened bugs in all project}
\label{fig_RQ1_percentage}       
\end{figure}

\noindent\textbf{Step 4.2 Training the model.} We use the training set to train the reopened bug prediction model. In our analysis, we use 7 commonly used models in software analytics as mentioned in Table~\ref{table:hyperparameters}. We choose these models in particular as a recent study by~\cite{agrawal2019dodge} notes,  Decision Tree, Random Forest, Logistic Regression, Multinomial Naive Bayes and K-nearest neighbors (KNN) are five of the most commonly used models in software analytics. In addition, we also include Gradient Boosting and Adaboost models as they are also commonly used in several software analytics studies \citep{ghotra2015revisiting,tantithamthavorn2018impact,tantithamthavorn2016automated}.


\begin{table}[!ht]
\caption{Overview of the used hyperparameters in our study. The default values are shown in bold. Randuniform denotes drawing random samples from a uniform distribution, and intranduniform denotes denotes drawing random samples from a uniform distribution and converting them to their nearest integer value.}
\label{table:hyperparameters}

\begin{tabular}{|l|l|l|l|}
\hline
\textbf{Model}                                                & \textbf{\begin{tabular}[c]{@{}l@{}}Parameter\\ name\end{tabular}} & \textbf{Parameter description}                                                                                         & \textbf{Parameter range}                                                \\ \hline
\begin{tabular}[c]{@{}l@{}}Random\\ Forest\end{tabular}            & n\_estimators                                                     & \begin{tabular}[c]{@{}l@{}}The number of trees in\\ the forest\end{tabular}                                           & \begin{tabular}[c]{@{}l@{}}(\textbf{100},\\ intranduniform(50, 150))\end{tabular}   \\ \hline
                                                                   & criterion                                                         & \begin{tabular}[c]{@{}l@{}}The function for\\ measuring the split\\ quality\end{tabular}                              & (\textbf{gini}, entropy)                                                         \\ \hline
                                                                   & \begin{tabular}[c]{@{}l@{}}min\_samples\\ \_split\end{tabular}    & \begin{tabular}[c]{@{}l@{}}The minimum needed\\ samples to split a node\end{tabular}                                  & \begin{tabular}[c]{@{}l@{}}(\textbf{2},\\ randuniform(0.0, 1.0))\end{tabular}    \\ \hline
\begin{tabular}[c]{@{}l@{}}Gradient\\ Boosting\end{tabular}        & n\_estimators                                                     & \begin{tabular}[c]{@{}l@{}}The number of boosting\\ stages to perform\end{tabular}                                    & \begin{tabular}[c]{@{}l@{}}(\textbf{100}, \\ intranduniform(50, 150))\end{tabular}  \\ \hline
                                                                   & \begin{tabular}[c]{@{}l@{}}min\_samples\\ \_leaf\end{tabular}     & \begin{tabular}[c]{@{}l@{}}The minimum needed\\ samples to be at a leaf\\ node\end{tabular}                           & \begin{tabular}[c]{@{}l@{}}(\textbf{1}, \\ intranduniform(1, 10))\end{tabular}      \\ \hline
\begin{tabular}[c]{@{}l@{}}Logistic\\ Regression\end{tabular}      & penalty                                                           & \begin{tabular}[c]{@{}l@{}}To specify the used\\ norm in the penalization\end{tabular}                                & (l1,\textbf{l2})                                                                 \\ \hline
                                                                   & tol                                                               & \begin{tabular}[c]{@{}l@{}}The tolerance for the\\ stopping criteria\end{tabular}                                     & \begin{tabular}[c]{@{}l@{}}(\textbf{1e-4}, \\ randuniform(0.0,0.1))\end{tabular} \\ \hline
                                                                   & C                                                                 & \begin{tabular}[c]{@{}l@{}}The inverse of\\ regularization strength\end{tabular}                                      & \begin{tabular}[c]{@{}l@{}}(\textbf{1}, \\ intranduniform(1,500))\end{tabular}          \\ \hline
Adaboost                                                           & n\_estimators                                                     & \begin{tabular}[c]{@{}l@{}}The maximum number \\ of estimators at which \\ the boosting is \\ terminated\end{tabular} & \begin{tabular}[c]{@{}l@{}}(\textbf{50}, \\ intranduniform(50, 150))\end{tabular}   \\ \hline
\begin{tabular}[c]{@{}l@{}}Multinomial \\ Naive Bayes\end{tabular} & alpha                                                             & \begin{tabular}[c]{@{}l@{}}The additive smoothing\\ parameter\end{tabular}                                            & \begin{tabular}[c]{@{}l@{}}(\textbf{1}, \\ randuniform(0.0, 1.0))\end{tabular}       \\ \hline
\begin{tabular}[c]{@{}l@{}}Decision\\  Tree\end{tabular}           & splitter                                                          & \begin{tabular}[c]{@{}l@{}}Choosing strategy for\\ splitting at each node\end{tabular}                                & \begin{tabular}[c]{@{}l@{}}(\textbf{best}, random)\end{tabular}               \\ \hline
                                                                   & criterion                                                         & \begin{tabular}[c]{@{}l@{}}The function for measu-\\ ring the split quality\end{tabular}                              & \begin{tabular}[c]{@{}l@{}}(\textbf{gini},  entropy)\end{tabular}              \\ \hline
                                                                   & \begin{tabular}[c]{@{}l@{}}min\_samples\\ \_split\end{tabular}    & \begin{tabular}[c]{@{}l@{}}The minimum needed\\ samples to split a node\end{tabular}                                  & \begin{tabular}[c]{@{}l@{}}(\textbf{2}, \\ randuniform(0.0,1.0))\tablefootnote{\url{https://scikit-learn.org/stable/modules/generated/sklearn.tree.DecisionTreeClassifier.html}}\end{tabular}    \\ \hline
KNN                                                                & n\_neighbors                                                      & \begin{tabular}[c]{@{}l@{}}The number of neigh-\\ bors\end{tabular}                                                   & \begin{tabular}[c]{@{}l@{}}(\textbf{5}, \\ intranduniform(2, 25))\end{tabular}          \\ \hline
                                                                   & weights                                                           & \begin{tabular}[c]{@{}l@{}}The weight function\\ used in prediction\end{tabular}                                      & \begin{tabular}[c]{@{}l@{}}(\textbf{uniform}, distance)\end{tabular}          \\ \hline
                                                                   & p                                                                 & Power parameter                                                                                                       & \begin{tabular}[c]{@{}l@{}}(\textbf{2}, \\ intranduniform(1,15))\end{tabular}           \\ \hline
                                                                   & metric                                                            & \begin{tabular}[c]{@{}l@{}}The distance metric to\\ use for the tree\end{tabular}                                     & \begin{tabular}[c]{@{}l@{}}(\textbf{minkowski}, \\ chebyshev)\end{tabular}       \\ \hline
\end{tabular}
\end{table}


\noindent\textbf{Step 4.3 Tuning the hyperparameters of the model.} 
We use a grid search \citep{sklearn_website_grid_search} based hyperparameters tuning to optimize the performance of the model. Grid search exhaustively considers all combinations of parameter values used to train the model and chooses the parameter that yields the best performance (we use AUC) for the constructed reopened bug prediction model.
Shihab et al. \citep{shihab2013studying} used a Decision Tree model without tuning the hyperparameters to build their reopened bug prediction model. However, several recent studies have shown that tuning the hyperparameters of a model is pivotal to ensure its optimal performance and interpretation~\citep{fu2016tuning,tantithamthavorn2018impact,tantithamthavorn2016automated}.


\noindent\textbf{Step 5: Model performance evaluation.} In this step, we measure the performance of various models on determining the likelihood of reopening a bug. We choose the AUC measure because the AUC measure is both threshold-independent and class imbalance insensitive. Several prior studies recommend the usage of AUC performance measure to evaluate the performance of the models in software analytics studies \citep{tantithamthavorn2016automated, rajbahadur2017impact, rajbahadur2019impact, ghotra2017large, lessmann2008benchmarking}. We find the performance of the reopened bug prediction model on a given project to be acceptable if on the given project mean AUC value of the constructed reopened bug prediction model across the 100 iterations is greater than 0.7 \citep{jiarpakdee2020impact, morasca2020assessment, al2014predicting, turabieh2019iterated}. Otherwise, we deem the performance of the reopened bug prediction model as poor.

\noindent\textbf{Step 6: Ranking important features.} 
To generate the feature importance in predicting the reopened bugs, we use the permutation feature importance method \citep{altmann2010permutation}. We use the permutation feature importance method over the top node analysis used by Shihab et al. \citep{shihab2013studying} for the following reasons. First, Shihab~et~al. compute the feature importance ranks on re-balanced datasets. Second, the results of top node analysis can be biased since Decision Tree favors categorical features \citep{altmann2010permutation}. To avoid these shortcomings, we use a permutation feature importance method. The permutation feature importance method works as follows. First, we compute the performance of reopened bug prediction model. We then randomly permute the values of each feature at a time and compute the model's performance to note how much performance drop does permuting a feature encounters when compared to the model built on all the features whose values are not permuted. A larger drop in performance due to a permutation of a feature's value signifies the higher importance of that feature. 

For computing the feature importance ranks, we use the dataset as it is to train the model (i.e., we do not use SMOTE in feature importance analysis). \cite{tantithamthavorn2018impact} find that using class re-balancing techniques when computing feature importance ranks results may lead to wrong features being considered as the important features. We compute the feature importance ranks only on the projects on which the best performing reopened bug prediction models have an acceptable AUC as~\cite{lipton2018mythos,chen2018applications} argue that for a model to be interpreted, it needs to have an acceptable performance. After this step for each iteration of training data, feature importance scores are obtained for each feature. More details of the working of permutation feature importance method can be found in \cite{altmann2010permutation}. After generating the feature importance score, we compute the features importance ranks using Scott-Knott ESD test \citep{tantithamthavorn2016empirical}. These ranks determine the order of importance of features in predicting the reopened bugs. We finally note the most common top two most important features used by the best performing model across all the studied projects.

\end{quote}

\noindent \textbf{Results:} \textbf{Across the 47 studied projects, at best, we observe an acceptable AUC ($\geqslant$ 0.7) for 34\% (16/47) of the projects.} Table~\ref{tabel:percentage_good_performance} shows the percentage of projects on which our constructed reopened bug prediction models deliver an acceptable AUC. We observe that on 66\% projects we get a poor AUC (\textless{} 0.7) for predicting reopened bugs. In the study by Shihab et al. \citep{shihab2013studying}, they found that the prediction models have a very high performance within the three studied projects. However, our finding suggests that the reopened bug prediction model does not necessarily show a very high performance on all projects when testing with a larger collection of projects.

\textbf{Reopened bug prediction model pipeline constructed with Random Forest and Gradient Boosting models yield an acceptable performance on more studied projects (AUC $\geqslant$ 0.7) than a pipeline with other models.} From Table~\ref{tabel:percentage_good_performance}, we observe that the Random Forest model and Gradient Boosting model yield an acceptable AUC on 34\% of the studied projects. Whereas, the Decision Tree and KNN models give an acceptable AUC on only 10\% of the projects. Such a result argues that future studies should consider Random Forest and Gradient Boosting models to construct reopened bug prediction models.





\begin{table}[!ht]
\caption{Percentage of projects with which the constructed reopened bug prediction model pipeline with different models yields an acceptable AUC}
\label{tabel:percentage_good_performance}
\begin{tabular}{|l|r|r|r|}
\hline
\textbf{Model}                                                         & \multicolumn{1}{l|}{\begin{tabular}[c]{@{}l@{}}\textbf{Projects with} \\ \textbf{acceptable AUC} \\ \textbf{(AUC$\geqslant$ 0.7)} \\ \textbf{count (mean AUC)}\end{tabular}} & \multicolumn{1}{l|}{\begin{tabular}[c]{@{}l@{}}\textbf{Projects with}\\ \textbf{poor AUC} \\ \textbf{(i.e., AUC\textless{} 0.7)} \\ \textbf{count (mean AUC)}\end{tabular}} & \multicolumn{1}{l|}{\begin{tabular}[c]{@{}l@{}}\textbf{Percentage of} \\ \textbf{the acceptable} \\ \textbf{projects}\end{tabular}} \\ \hline
Random Forest                                                      & 16 (0.78)                                                                                                                         & 31 (0.63)                                                                                                               & \textbf{34\%}                                                                                          \\ \hline
Gradient Boosting                                                  & 16 (0.77)                                                                                                                         & 31 (0.63)                                                                                                               & \textbf{34\%}                                                                                          \\ \hline
Logistic Regression                                                & 14 (0.78)                                                                                                                         & 33 (0.63)                                                                                                               & 29.8\%                                                                                                 \\ \hline
Adaboost                                                           & 12 (0.79)                                                                                                                         & 35 (0.63)                                                                                                               & 25.5\%                                                                                                 \\ \hline
\begin{tabular}[c]{@{}l@{}}Multinomial \\ Naive Bayes\end{tabular} & 8 (0.75)                                                                                                                          & 39 (0.61)                                                                                                               & 17\%                                                                                                   \\ \hline
Decision Tree                                                      & 5 (0.82)                                                                                                                          & 42 (0.59)                                                                                                               & 10.6\%                                                                                                 \\ \hline
KNN                                                                & 5 (0.76)                                                                                                                          & 42 (0.58)                                                                                                               & 10.6\%                                                                                                 \\ \hline
\end{tabular}
\end{table}

\textbf{For the projects with an acceptable AUC, the \textit{number of comments} is the most important feature. Such a result is in contrast with the findings by Shihab et al. \citep{shihab2013studying} who found that \textit{comment text} is the most important feature in determining the likelihood of reopening a bug.} We observe that using the best performing model (i.e., Random Forest), the \textit{number of comments} is the most important feature to predict reopened bugs in  31\% of the projects with an acceptable AUC. We further find that the \textit{comment text} which Shihab~et~al. found to be the most important feature in 66\% (i.e., two out of three) projects, is the most important feature in only 12\% projects and second most important feature in 31\% projects. Future studies can leverage the number of comments as a parameter in selection criteria for reopened bugs study.

\begin{summary_new}[left=8pt,right=8pt,top=10pt,bottom=10pt,boxed title style={colback=summary_color}]{\bfseries{Summary of RQ1}}
Using our updated reopened bug prediction model pipeline onto 47 JIRA projects, 
we find that only on 34\% of the studied projects we are able to achieve an acceptable AUC ($\geqslant$ 0.7) for our constructed reopened bug prediction model. Furthermore, we observe that the Random Forest model and Gradient Boosting model give the best AUC in determining the likelihood of reopening a bug. Finally, we also observe that \textit{number of comments} is the most important feature in determining the likelihood of reopening a bug.

\end{summary_new}


\subsection{RQ2: Why do bugs get reopened?}
\label{RQ3}
\begin{figure}[!ht]
\centering
  \includegraphics[width=0.9\textwidth]{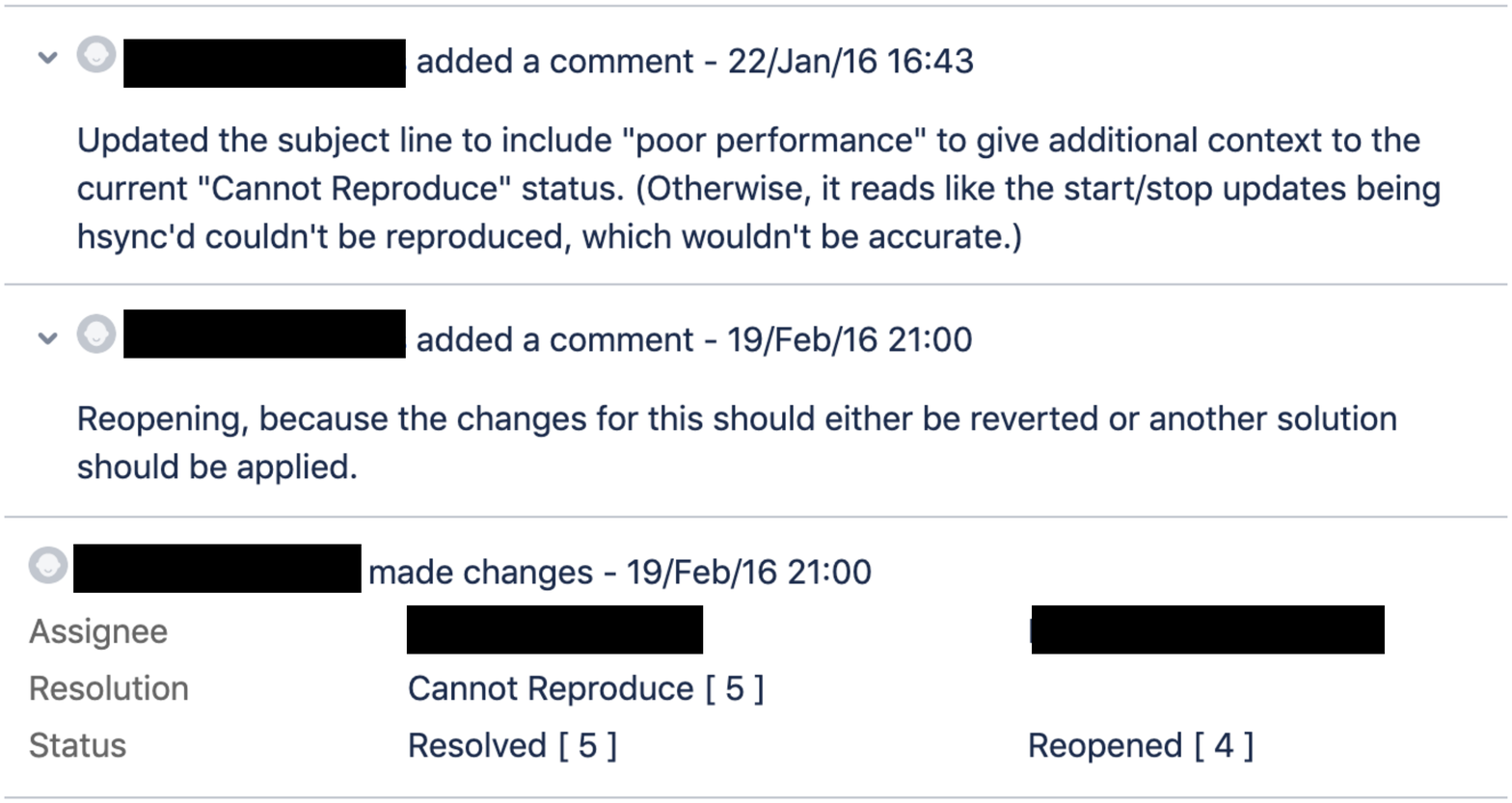}
\caption[Caption for LOF]{A bug report (ID: 4112) for the Apache Accumulo project, showing a comment depicting reason for reopening the bug at the same time when the bug is reopened, hence this comment is not used in reopened bug prediction model.}
\label{fig_RQ3_motivation}       
\end{figure}

\textbf{Motivation:} The reopened bug prediction model interpretation generates coarse insights in prior studies. For example, by interpreting the reopened bug prediction model, we could determine that \textit{number of comments} and \textit{comment text} are the top two most important features in determining the likelihood of reopening a bug; however, such insights do not provide any actionable suggestions that can be used to either prevent bug reopening or identify the root cause of reopening bugs. Moreover, we observe that some developers pointed out the rationale to reopen the bug during reopening the bug. For example, Figure \ref{fig_RQ3_motivation} shows a bug report\footnote{\url{https://issues.apache.org/jira/browse/ACCUMULO-4112}} (ID: 4112) for the Apache Accumulo project. In this bug report, the bug was considered as non-reproducible initially and resolved but later a developer reopened the bug and at the same time commented about the reason to reopen the bug. However, comments which are provided either during bug reopening or after reopening a bug are not used in the reopened bug prediction model in order to prevent data leak. This useful data (i.e., events during and post reopening bugs) can be analyzed using other approaches (such as a qualitative analysis) to generate insights into reopened bugs. Therefore, in this RQ, we investigate what are the rationale to reopen bugs. Moreover, since we observe in Section \ref{RQ2} that only 34\% of the studied projects give an acceptable AUC for predicting reopened bugs, in this RQ, we also investigate if the reasons for reopening a bug differ between the projects on which the constructed reopened bug prediction models have an acceptable and a poor performance. In order to better understand the rationale for reopening bugs, we leverage the whole history of bug reports to understand the characteristics of reopened bugs. We wish to gain insights into the challenges in building reopened bug prediction models for better understanding and managing reopened bugs.\\

\noindent \textbf{Approach:} We used all the 9,993 reopened bugs from 47 projects in our analysis. A mixed methods approach was used (including both the quantitative and qualitative study approach) in our study. We initially performed a quantitative analysis to identify the co-occurring changes at the time of reopening the bug. We then performed a qualitative analysis to identify the reasons why bugs get reopened.\\

\noindent\textit{\textbf{Quantitative study:}} The aim of the quantitative study is to identify what are all the changes that occur in reopened bugs. For our quantitative study, we divided all the changes that occur in the bug reports into three categories (i.e., before reopening, during reopening, after reopening). We considered all such events that occur before reopening a bug as the \textit{before reopening} category and all such events that occur at the time of reopening a bug as the \textit{during reopening} category. We considered all such events that occur after a bug is reopened as the \textit{after reopening} category. We extracted all the fields of changes during reopening a bug and identify the commonly occurring change patterns.\\

\noindent\textit{\textbf{Qualitative study:}} For our qualitative study, we first randomly selected reopened bugs, we then conducted an open coding study \citep{corbin1990grounded} to understand the reasons that bugs get reopened.

There are 9,993 reopened bugs from the 47 projects on JIRA. Out of these bugs, we selected a random sample of 370 reopened bugs. Our selection reaches a statistically representative sample of all the reopened bugs associated with the 47 projects with a 95\% confidence level and a 5\% confidence interval \citep{boslaugh2012statistics,zhang2019empirical}. For our qualitative study, the first three authors of the paper examined all the comments associated with our manually studied bugs, i.e., comments before/during/after a bug is reopened. For example, in a bug report\footnote{\url{https://issues.apache.org/jira/browse/FLEX-34059}} (ID: 34059) for the Apache Flex project, a developer commented in 40 minutes after the bug was reopened that: ``\textit{because the issue also happens with TextInputSkin (see ML piotr.zarzycki )}.''

Three coders (i.e., the first three authors) conducted the manual study of tagging the reasons to reopen bugs. We aimed to identify two labels, i.e., what is the reason to reopen a bug, and whether the rationale for reopening the bug occurred before, during, or after the bug is reopened? We discuss the steps in the open coding process below:

\begin{quote}

\noindent \textbf{Step 1: Round-1 coding.} We randomly selected 50 bugs out of the 370 reopened bugs. We tagged these bugs individually. Each coder's tagging process was independent. We did not interfere in other coder's tagging process.

\noindent \textbf{Step 2: Discussion after Round-1 coding.} After the first round, we obtained around 23 tags. These tags are specific reasons to reopen a bug. We then discussed to cluster these tags into categories of tags. After this meeting, we finalized the tag categories for the Round-1 coding.

\noindent \textbf{Step 3: Revisit Round-1 coding.} After the discussion, we re-coded the same 50 bugs with the updated tag categories.

\noindent \textbf{Step 4: Round-2 coding.} The remaining 320 bugs were used for the Round-2 coding. We assigned each coder with two-thirds of the remaining bugs (i.e., 214 bugs), such that each bug was tagged exactly by two coders. We were allowed to add new tags if a bug could not be classified by the existing Round-1 tag categories. Round-2 coding was also done individually by each coder without the influence of other coders. It took eight hours for each one of the coders to finish Round-2 coding. 

\noindent \textbf{Step 5: Discussion after Round-2 coding.} After the Round-2 coding, we discussed the results and resolved any disagreement. After this meeting, we finalized the tags for our study.

\noindent \textbf{Step 6: Revisiting Round-1 and Round-2 coding.} After the discussion in the previous round, we finally revisited all the bugs of Round-1 and Round-2 coding to update the new tags obtained after Round-2 coding. This step took several hours for each coder. This step was performed without the interference of other coders so as to maintain the independence of decision while tagging bugs.

\noindent \textbf{Step 7: Computing inter-coder agreement.} We used the tags to compute the inter-coder agreement. The inter-coder agreement is a measure to compute the agreement reached in tagging a bug. A bug reopening reason is considered as a match if both the coders provide the same tag for a bug and a mismatch if both the coders provide different tags for a bug. We computed the inter-coder agreement using Krippendorff’s $\alpha$ \citep{artstein2008inter, hayes2007answering}. The inter-coder agreement for our study is 0.87 which concludes that our qualitative analysis is reliable as prior studies use $\alpha$ $\geqslant$ 0.8 as an indicator of reliable agreement~\citep{li2020qualitative,beckler2018reliability, webb2020evaluation, vassallo2020developers, scoccia2020web}.



\noindent \textbf{Step 8: Dealing with disagreements.} After coding all the bugs, we discussed the disagreements among the coders for bugs with different tags. If a bug was tagged differently by two coders, and they mutually did not agree to a common tag, then the tag of the bug was determined by the third coder.



\end{quote}

For understanding the rationale for reopening a bug, we used four predefined categories (i.e., before reopening, during reopening, after reopening, and no evidence found) as our tags. Each coder tagged the evidence for the reopened bugs that are assigned to him when labeling the reasons.

After completing the above qualitative study, we studied the distribution of reasons for reopening bugs in projects with an acceptable AUC and projects with a poor AUC. To determine the projects that have an acceptable AUC and a poor AUC in the reopened bug prediction model, we considered the findings from Section \ref{RQ2}. For this study, we classified the projects as acceptable and poor based on the best performing model (i.e., Random forest) identified in Section \ref{RQ2}. We labelled all the 370 manually studied reopened bugs based on the project category as \textit{reopened bug from projects with acceptable AUC} and \textit{reopened bug from projects with poor AUC}. After labeling these 370 manually studied reopened bugs, we observed that 80 reopened bugs have a project category tag as \textit{reopened bug from projects with acceptable AUC} and the remaining 290 reopened bugs have a project category tag as \textit{reopened bug from projects with poor AUC}. To compare the distribution of reasons to reopen a bug in projects with acceptable and poor AUC, we needed to use the same dataset as we used to train the reopened bug prediction model. In our reopened bug prediction model we only used the dataset before reopening the bug. Therefore, to compare the distributions of reasons to reopen bugs in projects with acceptable and poor AUC, we only used those bugs where the reason to reopen a bug was identified before reopening the bug. From our first qualitative study, we had obtained the evidence tags for reopened bugs with four predefined categories (i.e., before reopening, during reopening, after reopening, and no evidence found). Using these \textit{evidence} tags and \textit{project category} tags, we identified that 13 out of 80 reopened bugs with project category as \textit{reopened bug} have \textit{evidence} tag as \textit{before reopening}, and 44 out of 290 reopened bugs with project category as \textit{reopened bug} have \textit{evidence} tag as \textit{before reopening}.
Note that in this analysis, we only computed and trusted the insights from projects with an acceptable AUC. We did so, as prior work \citep{lipton2018mythos} states that insights from only projects with an acceptable performance can be used. Therefore initially, we considered only 13 reopened bugs in projects with acceptable AUC and 44 reopened bugs in projects with poor AUC. However, this sample of reopened bugs was too small for any meaningful conclusion from the study, therefore, we added more samples in each project category (i.e., \textit{reopened bug from acceptable project} and \textit{reopened bug from poor project}) of reopened bugs, until we obtained 50 reopened bugs with evidence tag as \textit{before reopening} from projects with acceptable AUC and 50 reopened bugs with \textit{evidence} tag as \textit{before reopening} from projects with poor AUC. \\

\noindent \textbf{Results: Reopened bugs have an average number of 2.4 other changes of the bug report during bug reopening. Apart from the obvious change of reverting the resolution (98\%) and posting a comment (63\%) at the time of reopening a bug, 12\% of our studied bugs are reassigned during reopening. From a qualitative study, we observe that in 77\% of the reopened bugs, the rationale for reopening is provided either during or after the reopening event to support why a bug was reopened. On the other hand, in 15\% of the reopened bugs, the rationale is provided before reopening, and in 7\% the rationale is not provided} Using our quantitative study, we identify all the field changes that occur in reopened bug reports. Table \ref{table:quantitative_study} shows the distribution of field changes in the reopened bugs in different time periods (i.e., before reopening, during reopening, and after reopening). While analyzing the reopened bugs, we observe that 19\% of the reopened bugs are reassigned at least once after being reopened. \cite{yadav2019ranking} observe that reassigning bugs during bug fixing is not always harmful since bug reassignment can involve knowledge transfer among developers. In our study, we observe that in 2\% (226 out of 9,993) of the reopened bugs, developers change the assignee of the bug and resolve/close the bug immediately, indicating that the bug was reopened solely for the purpose of updating the assignee field of the bug report; however, the bug was already fixed. For example, in a bug report\footnote{\url{https://issues.apache.org/jira/browse/AIRAVATA-741}} (ID: 741) for the Apache Airvata project, the bug assignee was changed while reopening the bug and the bug was immediately closed. In 9\% (990 out of 9,993) of the reopened bugs, the developers change the assignee and do not immediately close/resolve the bug, suggesting that the bug was not fixed and re-assigned to another developer during reopening. For example, in a bug report\footnote{\url{https://issues.apache.org/jira/browse/ACCUMULO-4640}} (ID: 4640) for the Apache Accumulo project, the bug was re-assigned at the time of reopening the bug; however, the new assignee worked on the bug later and eventually resolved the bug. Prior studies observe that with an increasing number of bug reassignments, time to fix the bug also increases \citep{guo2011not,saha2015understanding}. The bug resolution process (i.e., during bug reopening and after bug reopening) is rich and contains many insights about bug reopening activities; however, no prior study has used this dataset to generate insights into bug reopening.


\begin{table}[!ht]
\caption{The distribution of major field changes that occur in reopened bugs grouped by before, during, and after reopening a bug.}
\label{table:quantitative_study}
\begin{tabular}{|r|r|r|r|}
\hline
\textbf{Change field} & \textbf{Before} & \textbf{During} & \textbf{After} \\ \hline
Resolution     & 9,632 (96.4\%)   & 9,810 (98.2\%)   & 8,227 (82.3\%)  \\ \hline
Comment        & 9,371 (93.8\%)   & 6,301 (63.1\%)   & 8,666 (86.7\%)  \\ \hline
Assignee       & 4,956 (49.6\%)   & 1,216 (12.2\%)   & 1,916 (19.2\%)  \\ \hline
Fix version    & 5,053 (50.6\%)   & 1,072 (10.7\%)   & 4,962 (49.7\%)  \\ \hline
Attachment     & 3,750 (37.5\%)   & 211 (2.1\%)     & 2,104 (21.1\%)  \\ \hline
Link           & 2,144 (21.5\%)   & 199 (2\%)       & 1,796 (18\%)    \\ \hline
\end{tabular}
\end{table}

\textbf{71\% (i.e., 263 out of 370) of the manually studied reopened bugs, the rationale for reopening is provided at the time of reopening the bug.} Table \ref{Table_Reasons_and_count_RQ3} shows the rationale with their bug count and proportion. For example, in a bug report\footnote{\url{https://issues.apache.org/jira/browse/OFBIZ-2197}} (ID: 2197) for the Apache OFBiz project, the developer posted a comment about reopening the bug at the time of reopening the bug: ``\textit{Reopened issue to re-examine the latest patch of CSS changes to the bluelight theme.}'' In another example, a bug report\footnote{\url{https://issues.apache.org/jira/browse/IGNITE-11429}} (ID: 11429) for the Apache Ignite project was reopened to update the resolution from \textit{fixed} to \textit{duplicate}, which occurred at the same time as reopening the bug. In 6\% (i.e., 24 out of 370) of the studied reopened bugs, the rationale is provided after the bug is reopened. For example, in a bug report\footnote{\url{https://issues.apache.org/jira/browse/NETBEANS-3598}} (ID: 3598) for the Apache NetBeans project, after reopening the bug the developer discussed that the patch did not work.

Although we observed in Section \ref{RQ2}, that \textit{comment text} is among the most important features in determining the likelihood of reopening a bug, many rationale in the history of bug reports can't be leveraged in determining the likelihood of reopening a bug as they occur mainly at the time of bug reopening or even after a bug was reopened. However, we encourage future research to leverage this set of data (i.e., during and after bug reopening) to alleviate the reopened bug management.

\begin{table}[!ht]
\caption{Different reasons for reopening the bugs based on various rationale categories (i.e., before bug reopening, during bug reopening and after bug reopening}
\label{Table_Reasons_and_count_RQ3}
\begin{tabular}{|l|l|l|l|r|l|l|l|l|}
\hline
\textbf{Category}                                         & \textbf{Reason}                                                & \textbf{Explanation}                                                                                                           & \textbf{Example}                                                                                                                                   & \multicolumn{1}{l|}{\textbf{\begin{tabular}[c]{@{}l@{}}\# \\ proj-\\ ects\end{tabular}}} & \textbf{\begin{tabular}[c]{@{}l@{}}Overall \\ count\\  (\%)\end{tabular}} & \textbf{\begin{tabular}[c]{@{}l@{}}Before\\ reop-\\ ening\\ count\\ (\%)\end{tabular}} & \textbf{\begin{tabular}[c]{@{}l@{}}During\\ reop-\\ ening\\ count\\ (\%)\end{tabular}} & \textbf{\begin{tabular}[c]{@{}l@{}}After\\ reop-\\ ening\\ count \\ (\%)\end{tabular}} \\ \hline
\multirow{2}{*}{Technical}                                & Patch                                                          & \begin{tabular}[c]{@{}l@{}}Bug is not\\ fixed by \\ the patch.\end{tabular}                                                    & \begin{tabular}[c]{@{}l@{}}Reopening, sin-\\ ce it seems this\\ issue is causing\\ problems with\\ high load. Will\\ revert and test.\tablefootnote{\url{https://issues.apache.org/jira/browse/HBASE-14689}}\end{tabular} & \multirow{2}{*}{37}                                                                      & \multirow{2}{*}{\begin{tabular}[c]{@{}l@{}}202 \\ (54.6\%)\end{tabular}}  & \multirow{2}{*}{\begin{tabular}[c]{@{}l@{}}50 \\ (13.5\%)\end{tabular}}                & \multirow{2}{*}{\begin{tabular}[c]{@{}l@{}}142 \\ (38.4\%)\end{tabular}}               & \multirow{2}{*}{\begin{tabular}[c]{@{}l@{}}10 \\ (2.7\%)\end{tabular}}                 \\ \cline{2-4}
                                                          & Integration                                                    & \begin{tabular}[c]{@{}l@{}}Bug is reop-\\ ened because\\ of an issue\\ during the\\ patch integr-\\ ation process.\end{tabular} & \begin{tabular}[c]{@{}l@{}}This has been\\ reverted, it\\ breaks unit \\ tests.\tablefootnote{\url{https://issues.apache.org/jira/browse/AMBARI-14383}}\end{tabular}                                                       &                                                                                          &                                                                           &                                                                                        &                                                                                        &                                                                                        \\ \hline
\begin{tabular}[c]{@{}l@{}}Document-\\ ation\end{tabular} & Update                                                         & \begin{tabular}[c]{@{}l@{}}Bug is reop-\\ ened to upd-\\ ate a field.\end{tabular}                                             & \begin{tabular}[c]{@{}l@{}}reopen to\\ update version\tablefootnote{\url{https://issues.apache.org/jira/browse/HAWQ-227}}\end{tabular}                                                                                 & 31                                                                                       & \begin{tabular}[c]{@{}l@{}}90 \\ (24.3\%)\end{tabular}                    & \begin{tabular}[c]{@{}l@{}}3 \\ (0.8\%)\end{tabular}                                   & \begin{tabular}[c]{@{}l@{}}78 \\ (21.1\%)\end{tabular}                                 & \begin{tabular}[c]{@{}l@{}}9 \\ (2.4\%)\end{tabular}                                   \\ \hline
Human                                                     & \begin{tabular}[c]{@{}l@{}}Incorrect\\ assessment\end{tabular} & \begin{tabular}[c]{@{}l@{}}Bug is reop-\\ ened due to\\ incorrect \\ assessment.\end{tabular}                                  & \begin{tabular}[c]{@{}l@{}}not a duplicate.\\ Same stack \\ trace, but root\\ cause is \\ different\tablefootnote{\url{https://issues.apache.org/jira/browse/HADOOP-10589}}\end{tabular}                                   & 28                                                                                       & \begin{tabular}[c]{@{}l@{}}52 \\ (14.1\%)\end{tabular}                    & \begin{tabular}[c]{@{}l@{}}4 \\ (1.1\%)\end{tabular}                                   & \begin{tabular}[c]{@{}l@{}}43 \\ (11.6\%)\end{tabular}                                 & \begin{tabular}[c]{@{}l@{}}5 \\ (1.4\%)\end{tabular}                                   \\ \hline
No reason                                                 & No stated reason                                                      & \begin{tabular}[c]{@{}l@{}}No reason\\ was found in\\ the bug report\\ for reopening.\end{tabular}                             & Not A Problem\tablefootnote{\url{https://issues.apache.org/jira/browse/STORM-190}}                                                                                                                                      & 15                                                                                       & \begin{tabular}[c]{@{}l@{}}26 \\ (7\%)\end{tabular}                       & \multicolumn{1}{r|}{-}                                                                 & \multicolumn{1}{r|}{-}                                                                 & \multicolumn{1}{r|}{-}                                                                 \\ \hline
\textbf{Total}                                            &                                                                &                                                                                                                                &                                                                                                                                                    & 47                                                                                       & \begin{tabular}[c]{@{}l@{}}370 \\ (100\%)\end{tabular}                    & \begin{tabular}[c]{@{}l@{}}57 \\ (15.4\%)\end{tabular}                                 & \begin{tabular}[c]{@{}l@{}}263 \\ (71.1\%)\end{tabular}                                & \begin{tabular}[c]{@{}l@{}}24 \\ (6.5\%)\end{tabular}                                  \\ \hline
\end{tabular}
\end{table}

\textbf{We identify four categories of reasons for reopening bugs, including technical (i.e., patch/integration issues), documentation (bookkeeping), human, and no stated reason.} Table~\ref{Table_Reasons_and_count_RQ3} shows the reason categories together with their number/proportion, explanation, and examples. We discuss the reasons for reopening bugs below:

\begin{itemize}[wide = 0pt]
	\item \textbf{Inline with common intuition, the majority (i.e., 54\%) of our studied bugs are reopened due to technical issues during the bug fixing processes, that is, the failure of code patch and patch integration.} More specifically, 41\% (i.e., 154 out of 370) of the reopened bugs are due to patch issues. Some of the major patch issues include runtime errors, regression, code improvements, and code merge issues. For example, a bug report\footnote{\url{https://issues.apache.org/jira/browse/FLINK-5206}} (ID: 5206) for the Apache Flink project is reopened due to runtime exception. Patch issues can deprecate the quality of the system and hence need to be fixed before resolving the bug. However, the above bug was not carefully tested before it was marked as fixed. During bug reopening, a developer pointed out that there is a runtime exception issue in the system and it took around 1.3 years to fix the reopened bug. In addition, 13\% (i.e., 48 out of 370) of our studied reopened bugs are due to integration failure. Integration issues include build failures, tests failure, flaky tests\footnote{\url{https://whatis.techtarget.com/definition/flaky-test}}, and unit test failures. For example, in a bug report\footnote{\url{https://issues.apache.org/jira/browse/DAFFODIL-346}} (ID: 346) for the Apache Daffodil project, test cases failure leads to the bug reopening. In this bug, a developer had already warned to test the fix before closing the issue; however, after closing the bug, a developer found that test cases are failing leading to reopening the bug. It took three days extra to resolve the issue, from the time the test cases failure issue was first observed. \cite{stolberg2009enabling} suggested certain code integration testing techniques such as running all unit tests with every build and developing unit tests for all new code that allow fixing bugs at a cheaper cost. We suggest the developer community to follow such code integration testing techniques during bug fixing process in order to avoid reopening a bug later due to integration issues.
	
	
	
	\item \textbf{In 24\% (i.e., 90 out of 370) of the studied reopened bugs, bugs are reopened due to bookkeeping issues (i.e., updating a field in the bug report).} Bug reports have various bug fields, such as \textit{version}, \textit{assignee}, and \textit{resolution}. In order for developers to update these bug fields, the bug needs to be reopened. We observe that 10\% (i.e., 39 out of 370) bugs are reopened to update the fix version. For example, a bug report\footnote{\url{https://issues.apache.org/jira/browse/HAWQ-480}} (ID: 480) for the Apache HAWQ project was reopened just to update its associated version. Moreover, 7\% (i.e., 26 out of 370) bugs are reopened to update the resolution. For example, a bug report\footnote{\url{https://issues.apache.org/jira/browse/FELIX-1999}} (ID: 1999) for the Apache Felix project was reopened and the developer who reopened the bug posted a comment to point out that it was reopened to fix the resolution. 1\% (i.e., 5 out of 370) bugs are reopened to update the assignee. For example, in a bug report\footnote{\url{https://issues.apache.org/jira/browse/SPARK-3598}} (ID: 3598) for the Apache Spark project, the bug was reopened to change its assignee. Bugs that are reopened to update the meta information of a fixed bug can use other mechanisms to avoid any confusion with a reopened bug due to technical issues. So far, technical issues and documentation issues are not differentiated by prior studies in determining the likelihood of reopening a bug. New bug features can be used to better identify reopened bugs due to technical issues instead of documentation issues. 
	
	
	\item \textbf{In 14\% (i.e., 52 out of 370) of our studied reopened bugs, bugs are reopened due to an incorrect initial assessment.} These bugs include cases where a bug was wrongly assessed initially, leading to the reopening of the bug at a later stage. We observe that 3\% (i.e., 14 out of 370) of our studied bugs were reopened since they were wrongly classified as \textit{non-reproducible}. In this case, if a developer is not able to reproduce the bug, then he stops working on the bug and resolves the bug without fixing it by considering the bug as \textit{non-reproducible}. For example, in a bug report\footnote{\url{https://issues.apache.org/jira/browse/THRIFT-2570}} (ID: 2570) for the Apache Thrift project, a developer initially closed the bug by mentioning ``\textit{can't reproduce ... feel free to re-open if new information becomes available}'', but after almost a month, another developer reopened the bug and fixed the bug. To handle such issues, the reporter of the bug should provide detailed reproduction steps for assignees to work on the bug properly. 2\% (i.e., 10 out of 370) of our studied bugs were wrongly considered as a duplicate. For example, a bug\footnote{\url{https://issues.apache.org/jira/browse/SPARK-17024}} (ID: 17024) for the Apache Spark project is considered as a duplicate initially; however, after 8 days, another developer reopened the bug and started working on fixing the bug. Prior work \citep{jalbert2008automated} also studied duplicate bug reports and found that in some projects almost a quarter of all bugs are duplicate.
	
	\item \textbf{In 7\% (i.e., 26 out of 370) of our studied reopened bugs, we were not able to identify the reason for reopening the bug.} In these bugs, developers do not provide any rationale in the bug to identify the reason for reopening the bug. For example, in a bug report\footnote{\url{https://issues.apache.org/jira/browse/HAWQ-1152}} (ID: 1152) for the Apache HAWQ project, no comment is provided at the time of reopening the bug.

\end{itemize}

In addition, we compute the closed to reopened time and reopened to fixed time for all categories of reopened bugs, and compute whether these reopened bugs are finally fixed or not. Table \ref{Table_RQ3_Reasons_time_count} shows that for reopened bugs in the technical and human category, the median time from reopened to fixed is much longer than the median time from closed to reopened.

\begin{table}[!ht]
\caption{Distribution of the median time spent in various states of reopened bugs}
\label{Table_RQ3_Reasons_time_count}
\resizebox{\textwidth}{!}{%
\begin{tabular}{|
>{\columncolor[HTML]{FFFFFF}}l |
>{\columncolor[HTML]{FFFFFF}}l |
>{\columncolor[HTML]{FFFFFF}}l |
>{\columncolor[HTML]{FFFFFF}}l |
>{\columncolor[HTML]{FFFFFF}}l |}
\hline
{\color[HTML]{000000} \textbf{Category}} & {\color[HTML]{000000} \textbf{Reason}} & {\color[HTML]{000000} \textbf{\begin{tabular}[c]{@{}l@{}}\# of bugs\\ finally fixed\\ (\% bugs)\end{tabular}}} & {\color[HTML]{000000} \textbf{\begin{tabular}[c]{@{}l@{}}Median time\\ from closed\\ to reopened\\ (in days)\end{tabular}}} & {\color[HTML]{000000} \textbf{\begin{tabular}[c]{@{}l@{}}Median time\\ from reopened\\ to fixed\\ (in days)\end{tabular}}} \\ \hline
\cellcolor[HTML]{FFFFFF}{\color[HTML]{000000} } & {\color[HTML]{000000} Patch} & \cellcolor[HTML]{FFFFFF}{\color[HTML]{000000} } & \cellcolor[HTML]{FFFFFF}{\color[HTML]{000000} } & \cellcolor[HTML]{FFFFFF}{\color[HTML]{000000} } \\ \cline{2-2}
\multirow{-2}{*}{\cellcolor[HTML]{FFFFFF}{\color[HTML]{000000} Technical}} & {\color[HTML]{000000} Integration} & \multirow{-2}{*}{\cellcolor[HTML]{FFFFFF}{\color[HTML]{000000} 190 (94.1\%)}} & \multirow{-2}{*}{\cellcolor[HTML]{FFFFFF}{\color[HTML]{000000} 1.5}} & \multirow{-2}{*}{\cellcolor[HTML]{FFFFFF}{\color[HTML]{000000} 16}} \\ \hline
{\color[HTML]{000000} Documentation} & {\color[HTML]{000000} Update} & {\color[HTML]{000000} 87 (96.7\%)} & {\color[HTML]{000000} 2} & {\color[HTML]{000000} 0} \\ \hline
{\color[HTML]{000000} Human} & {\color[HTML]{000000} Incorrect assessment} & {\color[HTML]{000000} 47 (90.4\%)} & {\color[HTML]{000000} 0.7} & {\color[HTML]{000000} 17.7} \\ \hline
{\color[HTML]{000000} No reason} & {\color[HTML]{000000} No stated reason} & {\color[HTML]{000000} 26 (100\%)} & {\color[HTML]{000000} 0} & {\color[HTML]{000000} 0} \\ \hline
\end{tabular}%
}
\end{table}

\noindent\textbf{94\% and 4\% of the total bugs in projects with an acceptable AUC are reopened due to patch issues and integration issues respectively. However, in projects with a poor AUC, only 66\% and 22\% bugs are reopened due to patch issues and integration issues respectively.} Table \ref{Table:acceptable_vs_poor_before} shows the distribution of reason categories identified before reopening the bug together with their number/proportion for both projects with acceptable as well as poor AUC. In the projects where we get an acceptable AUC, we manually classify the bugs with reasons before the reopening event. We observe that in 50 out of 330 randomly studied reopened bugs from projects with acceptable AUC, the reason to reopen bugs was identified before bug reopening. Out of these 50 bugs, there are 49 (98\%) bugs where the reason to reopen a bug was identified before bug reopening are reopened due to technical issues; however, no bug is identified where the reason identified before bug reopening is due to documentation issues. For example, in a bug report\footnote{\url{https://issues.apache.org/jira/browse/CASSANDRA-2626}} (ID: 2626) for the Apache Cassandra project, the developer comments before reopening the bug that \textit{``Thanks for your quick response. I'm not sure but I think that the fix doesn't work. Types of pendingFlush, sstables and compacting objects are java.util.Collections.UnmodifiableSet. Those are not instances of org.apache.commons.collections.set.UnmodifiableSet.''} stating that the bug was reopened due to technical reasons. In projects with an acceptable AUC, a technical reopening reason can be identified before reopening events, suggesting that new features that can characterize technical issues can be used in a reopened bug prediction model. Moreover, we observe that in 50 out of 313 randomly studied reopened bugs from projects with poor AUC reason to reopen a bug was identified before bug reopening. Out of these 50 bugs, 11 (22\%) are reopened due to integration issues, this is higher than the amount of bugs reopened due to integration issues (i.e., 4\%) identified before bug reopening in projects with an acceptable performance. Apart from bugs reopening in projects with poor performance due to technical issues, 3 (6\%) bugs where the reason to reopen a bug was identified before bug reopening are reopened due to documentation issues. For example, in a bug report\footnote{\url{https://issues.apache.org/jira/browse/SPARK-15519}} (ID: 15519) for the Apache Spark project, a developer comments \textit{``I misclicked and resolved as Fixed instead of as Duplicate. Feel free to edit.''} Since bug reopening due to documentation issues is a type of trivial bookkeeping activity, we encourage future studies to remove the bug reopens due to documentation issues from their study of determining the likelihood of reopening a bug.


From the study using the before reopening bug dataset, we observe that 98\% and 88\% of bugs in projects with an acceptable and a poor performance respectively are reopened due to technical issues. All these reasons are identified by developer comments, suggesting that \textit{comment text} is an important feature in the study of reopened bugs, which supports our finding in RQ1. From the study using during/after bug reopening dataset and before bug reopening dataset, we observe that in 77\% of the bugs the reason to reopen a bug is identified during/after reopening the bug, this is 5 times more than bug reopening reasons identified before (i.e., 15\%) bug reopening, indicating that 5 times more data can be leveraged while considering during/after bug reopening dataset in future bug reopening studies. Moreover, it also signifies that useful discussions (i.e., comments about bug reopening reasons) occur more frequently during/after bug reopening than before bug reopening.

\begin{table}[!ht]
\caption{Different reasons for reopening the bugs identified before bug reopening in various project categories (i.e., acceptable, and poor)}
\label{Table:acceptable_vs_poor_before}

\begin{tabular}{|l|l|l|r|r|r|r|}
\hline
\textbf{Category}                                         & \textbf{Reason}                                                 & \textbf{\begin{tabular}[c]{@{}l@{}}Eviden-\\ ce time\end{tabular}} & \multicolumn{1}{l|}{\textbf{\begin{tabular}[c]{@{}l@{}}Accep-\\ table per-\\formance\\ projects\end{tabular}}} & \multicolumn{1}{l|}{\textbf{\begin{tabular}[c]{@{}l@{}}\# Extra \\ reopened\\ bugs need-\\ ed in acce-\\ ptable\\ projects\end{tabular}}} & \multicolumn{1}{l|}{\textbf{\begin{tabular}[c]{@{}l@{}}Poor per-\\ formance\\ projects\end{tabular}}} & \multicolumn{1}{l|}{\textbf{\begin{tabular}[c]{@{}l@{}}\# Extra \\ reopened\\ bugs need-\\ ed in poor\\ projects\end{tabular}}} \\ \hline
\multirow{2}{*}{Technical}                                & Patch                                                           & before                                                             & 47(94\%)                                                                                        & 250                                                                                                                                       & 33 (66\%)                                                                              & 23                                                                                                                              \\ \cline{2-7} 
                                                          & Integration                                                     & before                                                             & 2 (4\%)                                                                                         & 250                                                                                                                                       & 11 (22\%)                                                                              & 23                                                                                                                              \\ \hline
\begin{tabular}[c]{@{}l@{}}Documen-\\ tation\end{tabular} & Update                                                          & before                                                             & 0 (0\%)                                                                                         & 250                                                                                                                                       & 3 (6\%)                                                                                & 23                                                                                                                              \\ \hline
Human                                                     & \begin{tabular}[c]{@{}l@{}}Incorrect \\ assessment\end{tabular} & before                                                             & 1 (2\%)                                                                                         & 250                                                                                                                                       & 3 (6\%)                                                                                & 23                                                                                                                              \\ \hline
No reason         &                                        \begin{tabular}[c]{@{}l@{}}No stated \\ reason\end{tabular}                                                       & before                                                             & -                                                                                               & 250                                                                                                                                       & -                                                                                      & 23                                                                                                                              \\ \hline
\end{tabular}
\end{table}

\begin{summary_new}[left=8pt,right=8pt,top=10pt,bottom=10pt,boxed title style={colback=summary_color}]{\bfseries{Summary of RQ2}}
In 77\% of the reopened bugs, the reason to reopen a bug can be identified either during or after a bug is reopened; however, this rich dataset (i.e., during and after bug reopening) is never studied specifically. 54\% of our manually studied reopened bugs are due to technical issues (i.e., the initial patching/integration process fails). 24\% of our manually studied reopened bugs are due to documentation issues (i.e., updating a field of the bug). 14\% of our manually studied reopened bugs are due to human related issues (i.e., incorrect assessment). In projects with an acceptable AUC, 94\%, and 4\% of the bugs where the reason to reopen a bug is identified before bug reopening are due to patch issues and integration issues respectively, while in projects with a poor AUC, apart from patch issues (i.e., 66\%), bugs are reopened due to integration issues (i.e., 22\%).
\end{summary_new}





\section{The implications of our findings}
\label{Discussion}
Table \ref{table:discussion_table} shows the findings of our study and their implications. We discuss implications specific to each interest group (i.e., developers, and researchers). Note that, we think that these implications can be of more interest to a particular target group.

\begin{table}[!ht]
\caption{Our major findings and their implications.}
\label{table:discussion_table}
\begin{tabular}{|p{1.4cm}|p{5.3cm}|p{6cm}|}
\hline
\textbf{Audience}            & \textbf{Findings}                                                                                                                                                                                        & \textbf{Implications}                                                                                                                                                                      \\ \hline
Developers                   & 13\% (i.e., 48 out of 370) of the manually studied bugs are reopened due to the integration issues and 22\% (11 out of 50) of the bugs in projects with poor AUC are reopened due to integration issues. & Developers should use code integration testing techniques such as running all unit tests with builds before resolving bugs in order to avoid bug reopening later due to integration issues. \\ \hline
\multirow{6}{*}{Researchers} & Same as above                     & Current techniques on integration testing mainly focus on software bugs in general. Future research can propose novel integration testing techniques specifically for handling the reopened bugs to avoid extra effort in fixing such bugs.                                                                      \\ \cline{2-3} 
                             & 63\% and 86\% bugs have comments during and after bug reopening respectively. In 77\% bugs the reason to reopen a bug is identified either during or after reopening the bug.                                                                                                                                   & Researchers should leverage data from during/after bug reopening to determine challenges in reopened bug management. 
                             
                             \\ \cline{2-3} 
                             & 24\% of the bugs are reopened to update the fields of the bug report.                                                                                                                                  & Researchers should drop the sub-dataset containing bug reports reopened due to documentation (i.e., bookkeeping) issues. 
                             
                             \\ \cline{2-3} 
                             & Prior studies on the prediction of reopened bugs are based on dataset with data leak issues.                                                                                                                                   & Researchers should consider only those events that occur before a bug is reopened to predict reopened bugs. 
                             
                             \\ \cline{2-3} 
                             & Only 34\% studied projects give an acceptable performance on models trained to determine whether a bug will get reopened.                                                                                                                                  & Current reopened bug prediction models are able to perform well when the majority of issues are reopened due to patch issues and integration issues as we observe in Section \ref{RQ3}. However, when projects contain lots of reopened bugs due to non technical issues, the performance of reopened prediction models depreciates. Future research should consider better data filtering techniques when selecting projects to build reopened bug prediction models and there should be more research on how to build better reopened bug prediction models where reasons to reopen bugs are not primarily technical issues.
                             
                             \\ \cline{2-3} 
                             & The number of comments is the most important feature to predict reopened bugs.                                                                                                    & Future studies on reopened bug prediction can leverage the number of developer comments as a factor to select reopened bugs for their study.                                                                                \\ \hline
\end{tabular}
\end{table}

\subsection{Implications for developers:} 

\textbf{Developers should use code integration testing techniques such as running all unit tests with builds before resolving bugs in order to avoid bug reopening later due to integration issues.} We observe in RQ2 (Section \ref{RQ3}) that 13\% (i.e., 48 out of 370) of the manually studied reopened bugs are due to the integration issues. Moreover, we also observe in RQ2 that 22\% of the bugs in projects with poor AUC are reopened due to integration issues identified before bug reopening. These issues include build failure, tests failure, unit tests failure, flaky tests, and tests timed out. Some of these issues can be handled during the bug fixing process, such as by executing unit tests locally and only committing the patch when the unit tests pass. Developers can also use automated tools for verifying the test accuracy. \cite{stolberg2009enabling} suggested certain code integration testing techniques such as running all unit tests with every build and developing unit tests for all new code that allow fixing bugs at a cheaper cost. Developers can use such techniques to resolve the bugs in order to avoid bugs getting reopened due to integration issues. Other issues, particularly the build success can also be tested after committing the patch. However, developers may be careless in performing such tests, or simply skip them to directly resolve the bugs. Additionally, we observe that there is a varying trend among development teams in waiting for the build to succeed and then resolving bugs. For example, in a bug report\footnote{\url{https://issues.apache.org/jira/browse/THRIFT-4531}} (ID: 4531) for the Apache Thrift project, the developer waits for the build to succeed then resolves the bug.
Whereas, in a bug report\footnote{\url{https://issues.apache.org/jira/browse/HBASE-15977}} (ID: 15977) for the Apache HBase project, the developer resolves the bug immediately after committing the patch without waiting for the build to succeed, later the build fails leading to reopening the bug. To avoid build failures in the master branch, the developer can use a feature branch to test their patch before merging the patch to the master branch. For example, in a bug report\footnote{\url{https://issues.apache.org/jira/browse/GEODE-1716}} (ID: 1716) for the Apache Geode project, the developer tests the patch in feature branch, and later the bug is resolved. A better mechanism to control the quality of test/build processes in bug fixing is needed so that developers can avoid reopening bugs later on.

\subsection{Implications for the researchers:} 

\textbf{Researchers can propose novel integration testing techniques specifically for handling the reopened bugs to avoid extra effort in fixing such bugs.} We observe in RQ2 (Section \ref{RQ3}) that 13\% of the manually studied bugs are reopened due to integration issues. Current techniques on integration testing mainly focus on software bugs in general \citep{herzig2015empirically, rodriguez2020bugs}. However, no prior research specifically focuses on integration testing techniques for reopened bugs. Therefore, we invite future research to conduct an experimental study in designing integration tests for reopened bugs so that the fixing process of reopened bugs can be improved.

\textbf{Researchers should leverage data from during/after bug reopening to determine challenges in reopened bug management.} We observe in RQ2 (Section \ref{RQ3}) that, 63\% of bugs have comments during reopening and 86\% of bugs have comments after bug reopening. These during and after reopening comments which are not used in reopened bug prediction studies, can be used to understand developer discussions of reopened bugs in order to determine the cause of bug reopening and its remedy. For example, in a bug report\footnote{\url{https://issues.apache.org/jira/browse/ZEPPELIN-5011}} (ID: 5011) for the Apache Zepplin project, the bug is reopened and later the developer comments \textit{``The build worked after I changed -Pspark-3.0 to -Pspark-3.0.0. At the end I learnt that it had ignored -Pspark-3.0.0 as an invalid option.''} indicating the resolution of the reopened bug. In addition, we observe that 77\% of our manually studied bugs have reason to reopen the bug during or after reopening the bug. A lot of this untapped data can be used to understand the challenges faced by developers in reopened bugs.

\textbf{Researchers should drop the sub-dataset containing bug reports reopened due to documentation (i.e., bookkeeping) issues.} We observe in RQ2 that one of the reason to reopen bugs is documentation issues. For example, in a bug report\footnote{\url{https://issues.apache.org/jira/browse/HAWQ-889}} (ID: 889) for the Apache HAWQ project, the developer reopened the bug to update the fix version. Some bugs are reopened in order to change a trivial bug report field, whereas other bugs are reopened in order to rework on the failed attempt to fix a bug. Bugs that are reopened due to documentation issues can be exempt from the reopened bug prediction model in order for the prediction model to predict more relevant bug reopening activities.


\textbf{Researchers should consider only those events that occur before a bug is reopened to predict reopened bugs.} We observe that in prior studies \citep{shihab2013studying, xia2015automatic}, one out of three studied projects (i.e., Apache project) has bugs with the last status as \textit{reopened} to determine the likelihood of reopening a bug. This is a data leak issue, and the bugs with a data leak issue should not be considered in the prediction model. We suggest the research community to carefully use data in prediction models, especially in the case when events occur sequentially.

\textbf{Researchers should consider better data filtering techniques when selecting projects to build reopened bug prediction models and there should be more research on how to build better reopened bug prediction models where reasons to reopen bugs are not primarily technical issues.} We observe in Section \ref{RQ2} that only 34\% studied projects given an acceptable AUC on models trained to determine whether a bug will get reopened. Moreover, in Section \ref{RQ3} that 98\% of manually studied reopened bugs in projects with an acceptable AUC are due to patch and integration issues. However, in projects with poor AUC, the percentage of patch issues and integration issues drops by 10\%. We invite future research to consider better data filtering techniques when selecting projects for predicting reopened bugs. For example, we recommend future research to select projects with more percentage of technical issues (Patch and integration issues). We also invite researchers to generate efficient models that are tuned to perform well on predicting reopened bugs on projects where bugs are reopened primarily due to non-technical issues (such as documentation issues, human issues).

\textbf{Future studies on reopened bug prediction can leverage the number of developer comments as a factor to select reopened bugs for their study.} We observe in Section \ref{RQ2} that the \textit{number of comments} is the most important feature to predict reopened bugs. We suggest that researchers can leverage the number of developer comments and filter out reopened bugs with no/very less developer comments for their reopened bugs study.

\section{Related Work}
\label{RelatedWork}
\subsection{Bug Management}

Bug triaging refers to the activities performed by the developers on un-assigned bugs, such as determining whether the bug needs attention, if so assigning the bug to a developer and adding a milestone for the bug \citep{bortis2013porchlight}. \cite{xia2016improving} proposed a MTM (multi-feature topic model) for bug triaging. \cite{jalbert2008automated} proposed a system to automatically identify duplicate bugs. They observed that their system is able to reduce software development costs by removing 8\% duplicate bugs. \cite{tian2012improved} extended their work and improved the accuracy of duplicate bug detection. \cite{somasundaram2012automatic} recommended a list of components that are most relevant to a bug. \cite{xia2014empirical} found that most of the bugs set their priority and severity value as the default value. Prior studies also focussed on bug assignment tasks in the triaging process. \cite{murphy2004automatic} proposed machine learning algorithms using text categorization to predict assignee of the bug. Other studies by \citep{anvik2006should, xia2013accurate, tian2016learning} proposed models to recommend the assignee for a bug. Our study also investigate bug management with respect to how bugs are reopened. We extend prior studies of bug reopening to conduct a large scale study with a modern machine learning model pipeline to understand the generalizability of reopened bug prediction models.

\subsection{Predictive Models in Bug Management}
Prior studies leveraged machine learning models to predict various attributes in a bug report. \cite{shihab2010predicting} conducted an initial study to predict reopened bugs on the Eclipse dataset using four dimensions and 22 features. \cite{shihab2013studying} also extended their own work in another study to predict reopened bugs on three datasets (Apache, Eclipse, and OpenOffice). They built a Decision Tree model and performed the top node analysis to identify the most important features in predicting reopened bugs, and observed that \textit{comment text} is the most important feature in predicting reopened bugs in the Eclipse project. \cite{xia2013comparative} investigated the performance of various classifiers to predict reopened bugs. Among the 10 classifiers, they identified that Bagging and Decision Tree model gives the best performance. Xia et al. \citep{xia2015automatic} also showed that their results outperform the prior work. \cite{xia2015automatic} proposed the ReopenPredictor\footnote{https://github.com/xin-xia/reopenBug} tool to predict reopened bugs using the same dataset as used by \cite{shihab2013studying}. Xia et al. \citep{xia2015automatic} proposed a feature selection method for choosing the most substantial textual features (\textit{comment text} and \textit{description text}) for the imbalanced dataset from both reopened and non-reopened bugs. \cite{xuan2012developer} modeled the developer prioritization of the bugs to predict the reopening of a bug. \cite{xuan2014towards} discussed the benefits of data reduction techniques in reopened bug analysis. \cite{caglayan2012factors} studied bug activity dataset in a large-scale enterprise software product. They analyzed four dimensions (i.e., issue-code relation, issue proximity network, issue reports, and developer activity) as possible factors that may cause a bug to be reopened. \cite{zimmermann2012characterizing} conducted a study to find the root cause of reopened bugs and built a model to predict reopened bugs. \cite{tu2018careful} defined data leak as information from the future used in predicting events making the prediction models misleadingly optimistic. They examined the prior literature in prediction tasks related to bugs and found that 11 out of 58 studies have data leak issues.


Prior studies on reopened bug prediction models focused only on a small dataset (i.e., three projects). Moreover, we identified that the Apache project used in the prior studies has a data leak issue. Instead, in our study, we use 47 projects to predict reopened bugs. We also use 7 classifiers in our study with latest techniques for generating reopened bug prediction model pipeline. Moreover, we also conduct a qualitative study to deeply understand the reasons why bugs get reopened.




\subsection{Empirical Studies on Reopened Bugs}
The activities of reopening bugs are an indication of issues that may project as more effort to resolve. Prior studies empirically mined bug reopening to gain a deeper understanding of reopened bugs. \cite{shihab2013studying} conducted an investigation of the bug resolving time for reopened and non-reopened bugs in the Eclipse project. They observed that reopened bugs take at least twice as much time to resolve than non-reopened bugs (i.e., on average of 149 days for non-reopened bugs and 371 days for reopened bugs). \cite{guo2010characterizing} conducted an empirical study on reopened bugs in Microsoft Windows and observed that if a bug is reopened too many times then probably it will not be fixed. \cite{mi2018not} discussed a new type of reopened bug that has no direct impact on the user experience or the normal operation of the system. They showed statistically that such reopened bugs are significantly different from other reopened bugs. \cite{mi2016empirical} studied four open source projects, i.e., CDT, JDT, PDE, and Platform, to quantitatively analyze reopened bugs on their impacts, proportions, and evolution. They identified that 6\% to 10\% bugs are reopened and reopened bugs remain active between 1.6 and 2.1 times higher than non-reopened bugs. In our study, we observe more in-depth activities in the life cycle of reopened bugs. For example, we find that during bug reopening developers also change resolution (in 98\% cases) and post comments (in 63\% cases). Such activities can help identify the rationale for reopening a bug. We also observe that 12\% of our studied bugs are reassigned during reopening. Future studies can investigate effects of reopening a bug on bug reassignment.


\section{Threats to Validity}
\label{threats}
\subsection{External Validity}

We use the JIRA bug tracking system for our study of reopened bugs. There are other popular bug tracking systems that can be studied and we encourage future work to investigate the characteristics of reopened bugs in such systems. Our selection criteria involve the top 50 projects in JIRA based on the committers count. The committers count criteria may not be the only metric to indicate the popularity of projects. For example, some committers may not be active, leading to projects with highly active but less number of committers. Future work can leverage other criteria that can be used to select projects. The selection of our studied dataset has an impact on the model performance and the feature importance analysis findings in our study.

\subsection{Internal Validity}

We study the reasons why bugs get reopened in RQ2 (Section \ref{RQ3}). We select a sample of bugs for our qualitative analysis by examining 370 bug reports from 9,993 reopened bugs. Other bugs that are not part of our studied sample may contain reasons other than what we identified in our analysis. However, we minimize the sample bias by selecting a statistical representative sample with a 95\% confidence level and a 5\% confidence interval. In our study, we do not claim to find all the reasons to reopen a bug. To alleviate this threat, future research is encouraged to study a larger sample of reopened bugs.

\subsection{Construct Validity}

We considered 7 classifiers (i.e., Random Forest, Gradient Boosting, Logistic Regression, Adaboost, Multinomial Naive Bayes, Decision Tree, and KNN) to predict the reopened bugs. However, there are other classifiers that are not covered in determining the likelihood of reopening a bug. The selection of 7 classifiers for our study is motivated by the study conducted by \cite{agrawal2019dodge}. Agrawal et al. used five classifiers (i.e., Random Forest, Logistic Regression, Multinomial Naive Bayes, Decision Tree, and KNN) for their study, we also consider two more classifiers (i.e., Gradient Boosting, and Adaboost) in our study as they are extensively used in prior prediction studies \citep{ghotra2015revisiting,tantithamthavorn2018impact,tantithamthavorn2016automated}.

We use the AUC value of 0.7 as a threshold to determine projects with an acceptable performance in determining the likelihood of reopening a bug. Changing the threshold value can change the percentage of projects with an acceptable performance, and impact the relative performance of different classifiers. However, we observe that many prior studies consider 0.7 as an AUC threshold to be acceptable~\citep{jiarpakdee2020impact, morasca2020assessment, al2014predicting, turabieh2019iterated}.






\section{Conclusion}
\label{Conclusion}
A reopened bug can take considerably more time to fix, and developers can make a lot of effort in fixing reopened bugs. In this study, we revisit reopened bug prediction on a large dataset consisting of 47 projects tracked by JIRA and using an updated pipeline for reopened bug prediction. We share the findings of our paper: 1) Prior studies on reopened bug prediction are based only on three projects (i.e., Apache, Eclipse, and OpenOffice), and out of these projects, the Apache project has a data leak issue. Hence, prior studies on predicting reopened bugs are effectively based only on two projects. Moreover, these studies used old techniques for constructing a reopened bug prediction model pipeline, indicating that predicting a reopened bug is not a solved problem. 2) Using our updated reopened bug prediction model pipeline, we observe that only 34\% of our studied 47 projects have an acceptable performance (i.e., AUC $\geqslant$ 0.7) in determining the likelihood of reopening a bug. 3) For the projects where the reopened bug prediction does not perform well, we propose researchers to create new features representing technical (i.e., specifically patch related) and documentation (i.e., bookkeeping) issues and leverage them to predict reopened bugs.

Based on the findings in our study, we suggest the developer community to follow practices involving carefully examining the proposed patch and testing the build success before resolving a bug. We find that 36\% and 13\% of the reopened bugs don’t have explanations for reopening the bugs during and after reopening the bugs respectively. The bugs with bug reopening justifications are more likely due to technical issues, whereas  the bugs without reopening justification can be due to trivial issues such as documentation. We invite future work to investigate reasons to reopen these bugs by conducting a deeper analysis such as developer surveys. We also encourage future research to explore this new venue of reopened bugs, e.g., by proposing new explanatory models to understand reopened bugs from leveraging the whole history of bug report and exploring insights into fixing reopened bugs more efficiently.

\section*{Disclaimer}
Any opinions, findings, and conclusions, or recommendations expressed in this material are those of the author(s) and do not reflect the views of Huawei.

\bibliographystyle{spbasic_updated}      
\bibliography{sample_library}   

\newpage

\end{document}